\documentclass[useAMS,usenatbib]{mn2e}
\usepackage{txfonts}
\usepackage{graphicx}
\usepackage{natbib}

\def\del#1{{}}

\newcommand{\ltsima}{$\; \buildrel < \over \sim \;$}
\newcommand{\lsim}{\lower.5ex\hbox{\ltsima}}
\newcommand{\gtsima}{$\; \buildrel > \over \sim \;$}
\newcommand{\gsim}{\lower.5ex\hbox{\gtsima}}

\newcommand{\bra}{\langle}
\newcommand{\ket}{\rangle}
\newcommand{\dang}{d_\mathrm{A}}
\newcommand{\planck}{{\em Planck}}
\newcommand{\plancks}{{\em Planck}'s }

\newcommand{\nK}{{\em n}K}

\newcommand{\dd}{\mathrm{d}}
\newcommand{\e}{\mathrm{e}}

\title[All-sky thermal and kinetic SZ-maps for \planck]
{Detecting Sunyaev-Zel'dovich clusters with \planck:\\ I. Construction of all-sky thermal and kinetic SZ-maps}
\author[B. M. Sch\"afer, C. Pfrommer, M. Bartelmann, V. Springel and L. Hernquist]
{B. M. Sch\"afer$^{1}$\thanks{e-mail: spirou@mpa-garching.mpg.de (BMS);
pfrommer@mpa-garching.mpg.de (CP); mbartelmann@ita.uni-heidelberg.de (MB); volker@mpa-garching.mpg.de (VS);
lars@cfa.harvard.edu (LH)},
C. Pfrommer$^{1}$\footnotemark[1],
M. Bartelmann$^{2}$\footnotemark[1],
V. Springel$^{1}$\footnotemark[1] and
L. Hernquist$^{3}$\footnotemark[1] \\
$^1$Max-Planck-Institut f\"ur Astrophysik, Karl-Schwarzschild-Stra{\ss}e 1, Postfach 1317, 85741
Garching, Germany\\
$^2$Institut f\"ur theoretische Astrophysik, Tiergartenstra{\ss}e 15, 69121 Heidelberg, Germany\\
$^3$Harvard-Smithsonian Center for Astrophysics, 60 Garden Street, Cambridge, MA 02138, USA}

\begin{document}
\pagerange{\pageref{firstpage}--\pageref{lastpage}}
\pubyear{2003}
\maketitle
\label{firstpage}

\begin{abstract}
All-sky thermal and kinetic Sunyaev-Zel'dovich (SZ) maps are presented for assessing how well the \planck~mission can 
find and characterise clusters of galaxies, especially in the presence of primary anisotropies of the cosmic microwave 
background (CMB) and various galactic and ecliptic foregrounds. The maps have been constructed from numerical simulations of 
structure formation in a standard \mbox{$\Lambda$CDM} cosmology and contain all clusters out to redshifts of $z = 1.46$ with 
masses exceeding $5\times10^{13} M_{\sun} /h$. By construction, the maps properly account for the evolution of cosmic 
structure, the halo-halo correlation function, the evolving mass function, halo substructure and adiabatic gas physics. The 
velocities in the kinetic map correspond to the actual density environment at the cluster positions. We characterise the 
SZ-cluster sample by measuring the distribution of angular sizes, the integrated thermal and kinetic Comptonisations, the 
source counts in the three relevant {\em Planck}-channels, and give the angular power-spectra of the SZ-sky. While our results 
are broadly consistent with simple estimates based on scaling relations and spherically symmetric cluster models, some 
significant differences are seen which may affect the number of cluster detectable by \planck.
\end{abstract}


\section{Introduction}
The Sunyaev-Zel'dovich (SZ) effects \citep{1972SZorig,1980ARA&A..18..537S, 1993birkinshaw, 
1995ARA&A..33..541R} have evolved from physical peculiarities to valuable and sound observational tools in cosmology. 
The thermal SZ-effect arises because photons of the cosmic microwave background (CMB) experience Compton-collisions with 
electrons of the hot plasma inside clusters of galaxies and are spectrally redistributed. The amplitude of the 
modulation of the Planckian CMB spectrum is a measure of the cluster electron column density and temperature. 
Alternatively, CMB photons may gain energy by elastic Compton collisions with electrons of the intra-cluster medium (ICM) 
due to the peculiar motion of the cluster relative to the CMB. This so-called kinetic SZ-effect is proportional to  
the peculiar velocity weighted electron column density and directly measures the cluster's velocity component parallel to 
the line-of-sight relative to the comoving CMB frame.

The advancement in sensitivity and angular resolution of sub-millimeter and microwave receivers have allowed high 
quality interferometric imaging of more than fifty clusters of galaxies by ground based 
telescopes \citep{2002ARA&A..40..643C} out to redshifts of $\sim0.8$, despite incomplete coverage of the Fourier plane. 
Apart from its primary scientific objective, namely the cartography of the CMB with angular resolutions close to 
$5\arcmin$, the upcoming \planck~mission \citep{1995P&SS...43.1459M, 2000IAUS..201E...8T} will be an unique tool for observing 
clusters of galaxies by their SZ-signature. \planck~is expected to yield a cluster catalogue that is surpassing the classic 
optical Abell catalogues or any existing X-ray catalogue in numbers as well as in depth and sky coverage. 

The capability of \planck~to detect SZ-clusters has been the subject of many recent works, pursuing 
analytical \citep{1997A&A...325....9A, 2002hzcm.conf...81D, 2001A&A...370..754B, 2002MNRAS.335..984M} as well as 
semianalytical \citep{2001MNRAS.325..835K, 2003MNRAS.338..765H} and numerical approaches \citep{2002MNRAS.336.1057H, 
2003ApJ...597..650W}. Their consensus is an expected total number of a few times $10^4$ clusters and the detectability 
of (sufficiently massive) clusters out to redshifts of $z\lsim 1$. The authors differ mainly in the expected 
distribution of the detectable clusters in redshift $z$. Where adressed, the authors remain sceptic about the detectability 
of the kinetic SZ-effect. 

As a result of various approximations made, there are clearly limitations in these studies: Concerning the SZ profiles of 
isolated clusters, simplifying assumptions like spherical symmetry, complete ionisation and isothermality have usually 
been made. Analytical treatments mostly rely on $\beta$-profiles for modeling the spatial variation of the Compton-$y$ 
parameter. Temperature models mostly make use of scaling laws derived from spherical collapse theory or are taken from X-ray 
observations. Naturally, the halo-halo correlation function is not taken account of, neither do the velocities correspond to 
the actual density environment, they are commonly drawn from a (Gaussian) velocity distribution.

The primary application of the all-sky SZ-maps would lie in the assessment of the extent to which cluster substructure and 
deviations from spherical symmetry, the halo-halo clustering and deviations from the scaling-laws alter the predictions made 
based on analytic methods. 

Additionally, the investigations mentioned above lack the inclusion of galactic foregrounds 
\citep[for a comprehensive review of foregrounds concerning \planck, see][]{1999NewA....4..443B}, the 
thermal emission from planets and minor celestial bodies of the solar system, beam patterns and spatially non-uniform 
instrumental noise. In order to quantify the extent to which the galactic and ecliptic foregrounds impede the 
SZ-observations by \planck, i.e. down to which galactic latitudes clusters will be detectable, a detailed simulation 
is necessary. Furthermore, the noise patterns will be highly non-uniform due to \plancks scanning strategy. For 
investigating this issue, all-sky maps of the thermal and kinetic SZ-effects are essential.

In our map construction, we take advantage of two numerical simulations of cosmic structure formation: The Hubble-volume 
simulation, that provided a well-sampled cluster catalogue covering a large volume and secondly, a set of template clusters 
resulting from a gas-dynamical simulation on much smaller scales, allowing us to extract template clusters. For all clusters of 
the Hubble-volume simulation, a suitable template was chosen and after having performed a scaling operation to improve the 
match it is projected onto the celestial sphere at the position requested by the Hubble-volume catalogue. By construction, 
the resulting all-sky SZ-maps show halo-halo correlation even on large angular scales, incorporate the evolution of the mass 
function and have the correct size distribution. In the kinetic SZ-map, it is ensured that the cluster peculiar velocities 
correspond to the ambient cosmological density field. Furthermore, the template clusters do exhibit realistic levels of 
substructure and departures from isothermality, and their ensemble properties also account for scatter around the idealised 
scaling laws.Therefore, most of the imperfections of traditional approaches will be remedied by our map construction process. 
However, there are impediments that could not be disposed of: They include gas physics beyond adiabaticity, e.g. 
radiative cooling and supernova feedback, that significantly alter the baryon distribution and temperature profiles of the ICM 
and hence the SZ-amplitude, incomplete ionisation, inclusion of filamentary structures and uncollapsed objects or diffuse gas. 
Another process influencing the thermal history of the ICM is reionisation, which also had to be excluded. Yet another 
complication are non-thermal particle populations in clusters of galaxies that give rise to the relativistic SZ-effect 
\citep{1979ApJ...232..348W, 2000A&A...360..417E}.

The paper is structured as follows: After the definition of the basic SZ quantities in Sect.~\ref{sect_szdef}, the 
simulations are outlined in Sect.~\ref{sect_sim}. The construction of the maps is described in detail in 
Sect.~\ref{sect_map} and the properties of the resulting maps are compiled in Sect.~\ref{sect_results}. Finally, the 
conclusions are presented in Sect.~\ref{sect_summary}.

\section{Sunyaev-Zel'dovich definitions}\label{sect_szdef}
The thermal and kinetic Sunyaev-Zel'dovich effects are the most important sources of secondary anisotropies in the 
CMB. Compton interactions of CMB photons with electrons of the ionised ICM give rise to these effects and induce 
surface brightness fluctuations in the CMB sky, either because of the thermal motion of the ICM electrons (thermal SZ) or 
because of the bulk motion of the cluster itself (kinetic SZ).

The relative change $\Delta T/T$ in thermodynamic CMB temperature at position $\bmath{\theta}$ as a function of
dimensionless frequency $x=h\nu / (k_B T_\mathrm{CMB})$ due to the thermal SZ-effect is given by 
eqn.~(\ref{sz_temp_decr}):
\begin{eqnarray}
\frac{\Delta T}{T}(\bmath{\theta}) & = & y(\bmath{\theta})\,\left(x\frac{e^x+1}{e^x-1}-4\right)\mbox{ with }
\label{sz_temp_decr}\\
y(\bmath{\theta}) & = & 
\frac{\sigma_\mathrm{T} k_B}{m_\e c^2}\int\dd l\:n_\e(\bmath{\theta},l)T_\e(\bmath{\theta},l)\mbox{,}
\label{sz_temp_decr_y}
\end{eqnarray}
where the amplitude $y$ of the thermal SZ-effect is commonly known as the thermal Comptonisation parameter. It is 
proportional to the line-of-sight integral of the temperature weighted thermal electron density (c.f. 
eqn.~(\ref{sz_temp_decr_y})). $m_\e$, $c$, $k_B$ and $\sigma_T$ denote electron mass, speed of light, Boltzmann's 
constant and the Thompson cross section, respectively. The kinetic SZ-effect arises due to the motion of the cluster 
relative to the CMB rest frame parallel to the line of sight. The respective temperature change is given by:
\begin{equation}
\frac{\Delta T}{T}(\bmath{\theta}) = -w(\bmath{\theta})\mbox{ with }
w(\bmath{\theta}) = \frac{\sigma_\mathrm{T}}{c}\int\dd l\:n_\e(\bmath{\theta,l})\upsilon_r(\bmath{\theta},l)\mbox{.}
\label{sz_temp_decr_w}
\end{equation}
Here, $\upsilon_r$ is the radial component of the cluster velocity, i.e. the velocity component parallel to the 
line-of-sight. The convention is such that the CMB temperature is increased, if the cluster is moving towards the 
observer, i.e. if $\upsilon_r<0$. In analogy to $y$, the quantity $w$ is refered to as the kinetic Comptonisation parameter.

\section{Simulations}\label{sect_sim}
Due to the SZ-clusters being detectable out to very large redshifts, due to their clustering properties on very large angular 
scales, and due to the requirement of reducing cosmic variance when simulating all-sky observations as will be performed by 
\planck, there is the need for very large simulation boxes, encompassing look-back distances to redshifts of order $z\simeq1$ 
which corresponds to comoving scales exceeding 2~Gpc. Unfortunately, a simulation incorporating dark matter and gas dynamics 
that covers cosmological scales of that size down to cluster scales and possibly resolving cluster substructure is presently 
beyond computational feasibility. 

For that reason, a hybrid approach is pursued by combining results from two simulations: The Hubble-volume 
simulation \citep{2001MNRAS.321..372J, 2000MNRAS.319..209C}, and a smaller scale simulation including (adiabatic) gas physics 
\citep{2002ApJ...579...16W}. The analysis undertaken by \citet{2001A&A...370..754B} gives expected mass and redshift ranges 
for detectable thermal SZ-clusters, which are covered completely by the all-sky SZ-map presented here.

The assumed cosmological model is the standard \mbox{$\Lambda$CDM} cosmology, which has recently been supported by 
observations of the WMAP satellite \citep{2003astro.ph..2209S}. Parameter values have been chosen as $\Omega_\mathrm{M} 
= 0.3$, $\Omega_\Lambda =0.7$, $H_0 = 100\,h\,\mbox{km~}\mbox{s}^{-1}\mbox{ Mpc}^{-1}$ with $h = 0.7$, 
$\Omega_\mathrm{B} = 0.04$, $n_\mathrm{s} =1$ and $\sigma_8=0.9$.

\subsection{Hubble-volume simulation}
The Hubble-volume simulation is one of the largest simulations of cosmic structure formation carried out to date. The 
simulation domain is a box of comoving side length $3~\mathrm{Gpc}/h$ (for the standard \mbox{$\Lambda$CDM} cosmology) and 
comprises $10^9$ dark matter particles. The simulations used were carried out by the Virgo Supercomputing Consortium using 
computers based at the Computing Centre of the Max-Planck Society in Garching and at the Edinburgh parallel Computing 
Centre. The data are publicly available for download\footnote{{\tt http://www.mpa-garching.mpg.de/galform/virgo/hubble}}.

The light-cone output of the Hubble-volume simulation \citep{2002ApJ...573....7E} was used for compiling a cluster 
catalogue. This ensures, that the abundance of clusters at any given redshift $z$ corresponds to the level of advancement in 
structure formation up to this cosmic epoch. The minimal mass was set to $5\times10^{13}M_{\sun}/h$, which roughly corresponds  
to the transition mass between a rich group of galaxies and a cluster. In order to cover redshifts out to the anticipated limit 
for \planck, light-cone outputs of differing geometry were combined: First, a sphere covering the full solid angle of $4\pi$ 
was used for redshift radii of $z=0$ to $z=0.58$. For redshifts exceeding $z = 0.58$, the northern and southern octant data 
sets were added. The octant data sets span a solid angle of $\pi/2$ and were replicated by rotation in order to cover the full 
sphere. Table~\ref{table_virgo} summarises the properties of the different output geometries.  In this way, a cluster catalogue 
with cluster mass $M$, position on the sky $\bmath{\theta}$, redshift $z$ and peculiar velocity $\upsilon_r$ projected onto the 
line-of-sight was compiled, comprising a total number of 2035858 clusters. For the sky-map construction, the positions 
$\bmath{\theta}$ were interpreted as ecliptic coordinates, the default coordinate system for \planck.

\begin{table}
\vspace{-0.1cm}
\begin{center}
\begin{tabular}{ccccc}\hline\hline
\vphantom{\Large A}%
data set & $z$ range & $\Omega/\mathrm{sr}$ & $N_\mathrm{halo}$ & shape\\
\hline
\vphantom{\Large A}%
MS & $0.0\leq z\leq0.58$ & $4\pi$  & 564818 & sphere \\
NO & $0.58< z\leq1.46$   & $\pi/2$ & 182551 & northern octant shell\\
PO & $0.58< z\leq1.46$   & $\pi/2$ & 185209 & southern octant shell\\
\hline
\end{tabular}
\end{center}
\caption{Basic characteristics of the light-cone outputs used for compiling a cluster catalogue. The columns denote the 
label of the data set, the range in redshift $z$, the solid angle $\Omega$ covered, the number of objects 
$N_\mathrm{halo}$ retrieved, and the geometrical shape.}
\label{table_virgo}
\end{table}

Here, it should be mentioned that the combination of different outputs gives rise to boundary discontinuities, at the 
surface of the central sphere as well as on the faces of the octant shells. These discontinuities do not only show up 
in the spatial halo distribution, but also in the velocities of clusters close to simulation box boundaries. 
Furthermore, the cluster catalogues exhibit small completeness deficiencies close to the edges of the simulation domain.

\subsection{Small scale SPH cluster simulations}\label{sect_hydrosim}
A hydrodynamical simulation of cosmological structure formation \citep{2002ApJ...579...16W} constitutes the basis of the 
SZ-template map construction. The simulation was performed with the {\tt GADGET} 
code\footnote{{\tt http://www.mpa-garching.mpg.de/galform/gadget/index.shtml}} \citep{2001NewA....6...79S} using the 
`entropy-conserving' formulation of SPH \citep{2002MNRAS.333..649S}. The simulation, first analysed in
\citet{2002ApJ...579...16W}, followed $216^3$ dark matter particles as well as $216^3$ gas particles in a cubical box of 
comoving side length $100\mbox{ Mpc}/h$ with periodic boundary conditions.  Purely adiabatic gas physics and shock heating were 
included, but radiative cooling and star formation were ignored, which however does not result in significant differences in SZ 
morphology, as shown by \citet{2002ApJ...579...16W}, but would impact on the scaling relations as demonstrated by 
\citet{2001ApJ...561L..15D}. We analyse 30 output redshifts ranging from $z =0$ out to $z = 1.458$. The comoving spacing along 
the line-of-sight of two subsequent outputs is $100~\mbox{Mpc}/h$. Halos were identified using a friends-of-friends algorithm 
with linking length $b= 0.164$, which yields all member particles of cluster-sized groups. We then employed a spherical 
overdensity code to estimate the virial mass and radius of each cluster. We computed the mass $M_\mathrm{vir}$ inside a sphere 
of radius $r_\mathrm{vir}$, interior to which the average density was 200 times the critical density 
$\rho_\mathrm{crit}=3H(z)^2/(8\pi G)$. We imposed a lower mass threshold of $M_\mathrm{vir}\geq 5\times10^{13}M_{\sun}/h$ in 
order tomatch the lower mass limit adopted for the Hubble-volume cluster catalogue.

\section{Sunyaev-Zel'dovich map construction}\label{sect_map}
The construction of the all-sky SZ-map proceeds in three steps: First, a set of template cluster maps was derived based on
cluster data from a gas-dynamical simulation (Sect.~\ref{map_template}). Then, for each of the clusters in the 
cluster catalogue obtained from the Hubble-volume simulation, a suitable hydrodynamical cluster template has been selected, 
scaled in mass and temperature in order to better fit the cluster from the Hubble-volume catalogue (Sect.~\ref{map_scaling}), 
and, for the kinetic sky map, boosted to the radial peculiar velocity required by the Hubble-volume simulation. The 
last step is the projection onto a spherical celestial map (Sect.~\ref{map_project}). In the subsequent paragraph 
(Sect.~\ref{sect_szbg}), the completeness of the resulting SZ-maps is investigated analytically.

\subsection{SZ-template map preparation}\label{map_template}
Square maps of the Compton-$y$ parameter of the selected clusters were generated by SPH projection of all 
friends-of-friends identified member gas particles onto a Cartesian grid with $128^2$ mesh points. The (comoving) side 
length $s$ of the map was adapted to the cluster size, such that the comoving resolution $g=s/128$ of the grid is 
specific to a given map.

If the particle $p$ at position $\bmath{r}_p=\left(x_p,y_p,z_p\right)$ has a smoothing length $h_p$, an SPH electron
number density $n_p$, and an SPH electron temperature $T_p$, the Compton-$y$ parameter for the pixel at 
position $\bmath{x}$ is given by:
\begin{equation}
y(\bmath{x}) = \frac{\sigma_\mathrm{T} k_B}{m_e c^2}\frac{h_p^3}{g^2}
\sum_p\left[\:
\int\limits_{x-g/2}^{x+g/2}\!\!\!\dd x_p
\int\limits_{y-g/2}^{y+g/2}\!\!\!\dd y_p
\int\limits_{-h_p}^{h_p}\!\!\dd z_p\: \mathcal{K}\left(\frac{r}{h_p}\right) n_p T_p\right]
\label{eqn_yproj}
\end{equation}

\begin{equation}
\mbox{with }r = \sqrt{(x_p-x)^2+(y_p-y)^2+z_p^2}\mbox{.}
\end{equation}

Here, we assume complete ionisation and primordial element composition of the ICM for the determination of electron
number density and temperature. In this way, we produce projections along each of the three coordinate axes. The
function $\mathcal{K}$ is the spherically symmetric cubic spline kernel suggested by \citet{1985A&A...149..135M}, which 
was also used in the SPH simulation:
\begin{equation}
\mathcal{K}(u) = \frac{8}{\pi}\times
\left\{
\begin{array}
{l@{,\:}l}1-6u^2+6u^3 & 0 \leq u \leq 1/2\\2(1-u)^3 & 1/2 < u \leq 1 \\0 & u > 1
\end{array}
\right.
\mbox{ with } u = r / h_p\mbox{.}
\end{equation}

The fact that the kernel $\mathcal{K}$ has a compact support $u\in\left[0\ldots 1\right]$ greatly reduces the
computational effort. 

The kinetic maps were treated in complete analogy: Maps of the Thomson optical depth $\tau$ were derived by means of 
eqn.~(\ref{eqn_wproj}):
\begin{equation}
\tau(\bmath{x}) = \sigma_\mathrm{T}\frac{h_p^3}{g^2}
\sum_p\left[\:
\int\limits_{x-g/2}^{x+g/2}\!\!\!\dd x_p
\int\limits_{y-g/2}^{y+g/2}\!\!\!\dd y_p
\int\limits_{-h_p}^{h_p}\!\!\dd z_p\: \mathcal{K}\left(\frac{r}{h_p}\right)\, n_p \right]\mbox{.}
\label{eqn_wproj}
\end{equation}

In eqn.~(\ref{eqn_wproj}), the influence of velocity differences inside the clusters was omitted. At the stage of 
projecting the template clusters onto the spherical map, the $\tau$-map obtained is boosted with the peculiar 
line-of-sight velocity $\upsilon_r$ in units of the speed of light $c$ required by the entry in the Hubble-volume 
catalogue in order to yield a Compton-$w$ amplitude.

Neglecting velocity differences inside the clusters does not seriously affect the measurement of cluster peculiar 
velocities with the kinetic SZ-effect shown by \citet{2003ApJ...587..524N}, the scatter in the velocity estimates 
increases only little ($50-100\mbox{ km}/\mbox{s}$) when considering a rather narrow beam ($\Delta\theta=1\farcm0$ 
FWHM), while the kinetic SZ-amplitude remains an unbiased estimator of the peculiar velocity. For our application 
purpose, the situation is even less troublesome because of \plancks wide beams ($\gsim5\farcm0$ (FWHM)).

\begin{figure}
\resizebox{\hsize}{!}{\includegraphics{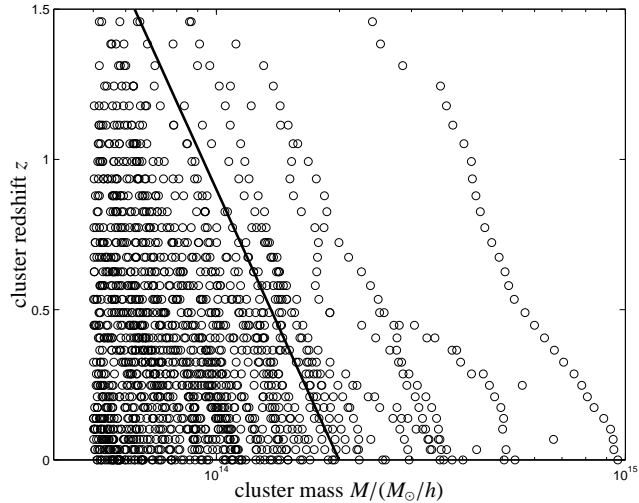}}
\caption{The population of our template clusters in the mass-redshift plane. The line separates the sparsely sampled 
region from the region in which a sufficient number of template clusters is available.}
\label{figure_mzplane}
\end{figure}

In this way, a sample of 1518 individual template clusters was obtained, and maps for projections along all three 
coordinate axes were derived, yielding a total of 4554 template maps for each of the two SZ-effects. 
Fig.~\ref{figure_mzplane} shows the distribution of clusters in the mass-redshift plane. Especially at high masses, the 
smooth growth of clusters by accretion can be clearly seen. Sudden jumps to larger masses are caused by the merging of 
low-mass clusters.

A limitation to our SZ-map construction is immediately apparent: The 4554 cluster template maps derived from the 
hydro-simulation are not strictly independent, but merely show the same clusters at different redshift. Thus, the morphological 
variety remains limited, but even though there is of course some variation in morphology due to accretion and merging events. 
This, however, may not be a severe restriction, keeping the wide \planck-beams in mind, that are unlikely to resolve cluster 
substructure for a large fraction of detectable clusters. In this case, the simulation will pick up mismatches in 
Comptonisation relative to the value expected from the spherical collapse model in conjunction with the Press-Schechter 
distribution of halo masses. 

\subsection{Cluster selection and scaling relations}\label{map_scaling}
In order to select a template map for projection, the closest template cluster in the $\log(M)$-$z$-plane for a given 
cluster from the Hubble-volume simulation was chosen. For the sparsely sampled region of the $M$-$z$-plane to the right of the 
line in Fig.~\ref{figure_mzplane}, a cluster from a pool containing the most massive clusters to the right of this line in the 
redshift bin under consideration was drawn. 

The template clusters are scaled in mass, temperature and spatial extent in order to yield a better match to the 
cluster from the Hubble-volume simulation according to formulae~\ref{eqn_mscaling}-\ref{eqn_tscaling}. The scaling 
is parameterised by the masses of the cluster of the Hubble-volume simulation $M_\mathrm{vir}^{(\mathrm{Hubble})}$ and 
the template cluster $M_\mathrm{vir}^{(\mathrm{template})}$:
\begin{eqnarray}
q_M & = & \frac{M_\mathrm{vir}^{(\mathrm{Hubble})}}{M_\mathrm{vir}^{(\mathrm{template})}} \label{eqn_mscaling}\\
q_R & = & \frac{r_1\,\left(M_\mathrm{vir}^{(\mathrm{Hubble})}\right)^{r_2} + r_3}
{r_\mathrm{vir}^{(\mathrm{template})}} \label{eqn_rscaling}\\
q_T & = &
\frac{t_1\,\left(M_\mathrm{vir}^{(\mathrm{Hubble})}\right)^{t_2} + t_3}
{t_1\,\left(M_\mathrm{vir}^{(\mathrm{template})}\right)^{t_2} + t_3} 
\label{eqn_tscaling}\mbox{.}
\end{eqnarray}

The parameters $r_i$ and $t_i$, $i\in\left\{1,2,3\right\}$, describing the scaling in size $q_R$ and in temperature 
$q_T$ were derived from template data: Fits to the virial radius as a function of mass and of the mean temperature 
inside the virial sphere as a function of mass were applied to the data of simulation outputs binned in five data sets. 
This approach leaves the map construction independent of idealised assumptions, like the prediction of cluster 
temperatures from the spherical collapse model, or from electron temperature measurements deduced from X-ray 
observations and keeps the weak trend of cluster temperature with redshift $z$ as contained in the simulations.

\begin{figure}
\resizebox{\hsize}{!}{\includegraphics{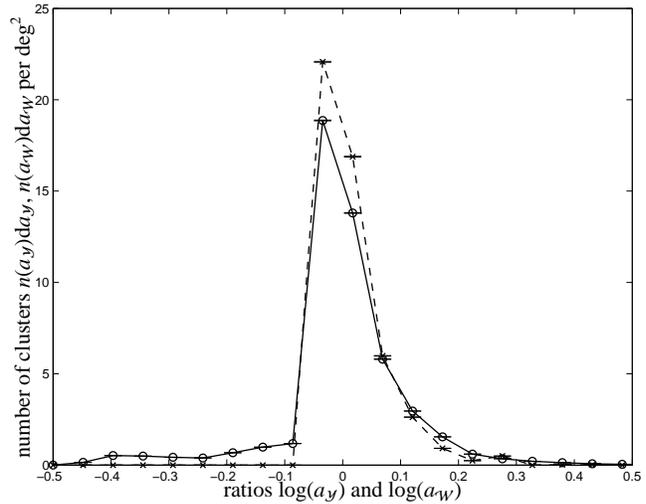}}
\caption{Distribution of the ratios $a_\mathcal{Y}$ (circles, solid line) and $a_\mathcal{W}$ (crosses, dashed line) 
that describes the impact of the mismatch between friends-of-friends masses and virial masses on the Comptonisations 
$\mathcal{Y}$ and $\mathcal{W}$.}
\label{fig_qmqt_distribution}
\end{figure}

Although the scaling has been constructed in order to yield the best possible match between the template cluster and the 
target cluster from the Hubble-volume simulation, there are artifacts in irregular systems due to inconsistencies in 
cluster masses $M_\mathrm{fof}^\mathrm{(template)}$ determined with a friends-of-friends algorithm for identifying 
cluster member particles and the virial mass estimates $M_\mathrm{vir}^\mathrm{(template)}$ following from applying the 
spherical overdensity code. After the scaling, these mismatches may be expressed as:

\begin{eqnarray}
a_\mathcal{Y} & = & q_M q_T M_\mathrm{fof}^\mathrm{(template)}/M_\mathrm{vir}^\mathrm{(Hubble)}\quad\mbox{and}\\
a_\mathcal{W} & = & q_M M_\mathrm{fof}^\mathrm{(template)}/M_\mathrm{vir}^\mathrm{(Hubble)}\mbox{.}
\end{eqnarray}

Fig.\ref{fig_qmqt_distribution} shows the distribution of ratios $a_\mathcal{Y}$ and $a_\mathcal{W}$ for the 
entire Hubble-volume catalogue. Clearly, one recognises large tails towards high values, because clusters have on 
average to be scaled to higher masses. This is due to the fact that the hydro-simulation outlined in 
Sect.~\ref{sect_hydrosim} does not sample the high-mass end of the Press-Schechter function satisfatorily, simply because of 
its small volume. Nevertheless, the mean of the distributions is close to one, which implies that the mismatches average out 
for the bulk of clusters.

\subsection{Projection onto the celestial sphere}\label{map_project}
For storing all-sky maps the HEALPix\footnote{{\tt http://www.eso.org/science/healpix/}} tesselation of 
the sphere proposed by \citet{2002adass..11..107G} has been chosen. In order to support structures as small as clusters, 
the $N_\mathrm{side}$ parameter has been set to 2048, resulting in a total number of $12~N_\mathrm{side}^2 = 50331648$ 
pixels. The side length of one pixel is then approximately $1\farcm71$, which is well below the anticipated Planck beam 
size of $5\farcm0$ in the highest frequency channels.

The scaled cluster maps are projected onto the spherical map by means of stereographic projection at the south ecliptic 
pole of the celestial sphere. By dividing the (comoving) position vector $(x,y)$ of a given pixel on the template map 
by the comoving angular diameter distance $\chi(z)$ at redshift $z$, one obtains the coordinates $(\alpha$,$\beta)$ on 
the tangential plane. Then, the stereographic projection formulae yield the (Cartesian) position vector 
$(\xi,\eta,\zeta)$ of this point projected onto the unit sphere:
\begin{eqnarray}
\bmath{r} = (\xi,\eta,\zeta+1) = 
\left(
\frac{4\alpha}{4+\alpha^2+\beta^2},\frac{4\beta}{4+\alpha^2+\beta^2},\frac{\alpha^2+\beta^2}{4+\alpha^2+\beta^2}
\right)\mbox{.}
\end{eqnarray}

In order to assign a Comptonisation amplitude to a given HEALPix pixel in the projection process, a solid angle 
weighted average is performed. For close-by clusters, the mesh size of the templates converted to angular units is 
larger than the HEALPix pixel scale. For those clusters, the map is refined iteratively by subdivision of a pixel into 4 
smaller pixels subtending a quarter of the original solid angle until the pixel size is well below the HEALPix pixel 
scale. Before projection, the template maps are smoothed with a Gaussian kernel with $\Delta\theta = 2\farcm0$, which is 
comparable to the HEALPix pixel scale. In this way, it is avoided that structures are destroyed by the combination of 
multiple template map pixels into a single HEALPix pixel. This convolution does not affect the later usage for 
simulations concering \planck: A second successive convolution with the narrowest beam results in an effective 
smoothing of $5.38\arcmin$, which corresponds to a decrease in angular resolution of roughly 7.5\%.

Additionally, a rotation of the template map around the $\bmath{e}_z$-axis about a random angle is performed in order 
to avoid spurious alignments of clusters. The projected pixels are then transported by Euler-rotations of the 
vector $(\xi,\eta,\zeta)$ to the position requested by the Hubble catalogue.

\subsection{Completeness of the all-sky SZ-maps}\label{sect_szbg}
The angular resolution of \planck~will not allow to spatially resolve low-mass and high-redshift clusters.  
There will be a Compton-$y$ background $\langle y_\rmn{bg}\rangle_\theta$ due to the higher number density of 
low-mass clusters compared to high-mass clusters which overcompensates their lower individual SZ-signature. Since 
ideally any isotropic background could be removed, we only have to take into account the average background fluctuation 
level $\langle y_\rmn{bg}^2\rangle_\theta$ which is described by power spectrum statistics. 

This section studies the influence of the background of unresolved SZ-clusters in our all-sky map of SZ-clusters on 
power spectrum statistics. Our simulation neglects the SZ signal of clusters both with masses smaller than $5 \times
10^{13} M_{\sun}/h$ and redshifts $z > 1.5$. In principle, these missing clusters could be accounted for by 
drawing them from a particular realisation of a {\em Poissonian random field} such that they obey the correct relative 
phase correlations, i.e.\ that they exhibit the observed local clustering properties. 

However, there are large uncertainties about the baryon fraction $f_\rmn{B}=\Omega_\rmn{B} /\Omega_\rmn{M}$ 
of low-mass halos($M_\rmn{halo} < 5 \times 10^{13} M_{\sun}$/h) especially at higher redshifts.  Analyses of X-ray observations 
of 45 local clusters ($z < 0.18$, only 4 of them lie at $z > 0.1$) carried out by \citet{1999ApJ...517..627M} and 
\citet{1999MNRAS.305..631A} suggest a weak trend of the cluster baryon fraction $f_\rmn{B}$ with cluster mass $M$ and a 
deviation from the universal value, which may be due to feedback processes like galactic winds that more effectively deplete 
the ICM of baryons in low-mass compared to high-mass clusters. The behaviour of $f_\rmn{B}$ at high redshifts is very 
uncertain, among other reasons because the relative importance of the different feedback processes at high redshift is yet 
unknown. This uncertainty is also reflected in different cooling rates and mechanisms, governing the ionisation fraction of the 
electrons and the resulting SZ flux of a particular cluster.
In the following, we study the contribution to the SZ flux of clusters both with masses smaller than $5 \times 
10^{13}\,M_{\sun}/h$ and redshifts $z >1.5$. Although the impact of this cluster population to the 
$\langle y_\rmn{bg}\rangle$-statistics amounts to a significant fraction, this population has a negligible contribution to the 
more relevant $\langle y_\rmn{bg}^2\rangle$-statistics which will be shown in the following.  The unresolved
cluster population is assumed to follow scaling relations derived from the spherical collapse model. Temperature $T$ and 
halo mass $M$ are assumed to be related by

\begin{equation}
  \label{eq:TMrelation}             
    \frac{k_B T}{6.03~\mbox{keV}} = \left(\frac{M}{10^{15} M_{\sun}/h}\right)^{2/3} (1+z) 
  \left( \frac{\Omega_0}{\Omega(z)}\right)^{1/3}
  \left( \frac{\Delta_\rmn{c}(z)}{178}\right)^{1/3}
\label{eqn_virial_temp}
\end{equation}
\citep[e.g. ][]{1996MNRAS.282..263E,1995MNRAS.275..720N}. The temperature $k_B T = 6.03 \mbox{ keV}$ for a cluster with 
$M = 10^{15} M_{\sun}/h$ has been adopted from \citet{2001ApJ...546..100M}. The density parameter at redshift $z$ is 
denoted by $\Omega(z)$, and $\Delta_\rmn{c}$ is the mean overdensity of a virialised sphere,
\begin{equation}
  \label{eq:overdensity}
  \Delta_\rmn{c} = 9 \pi^2\, \left\{1 + \alpha\left[\Omega(z)-1\right] + 
      \Omega(z)^\beta \right\}
\label{eqn_overdensity}
\end{equation}
with $(\alpha,\beta) = (0.7076,0.4403)$ for a flat cosmology
\citep{FelixStoehr1999}. Assuming that the total number $N_\e$ of thermal
electrons within the cluster virial radius is proportional to the virial mass yields
\begin{equation}
  \label{eq:Ne}
  N_\rmn{e} = \frac{1+f_\rmn{H}}{2} f_\rmn{B} \frac{M}{m_\rmn{p}},
\end{equation}
where $f_\rmn{H}$ is the hydrogen fraction of the baryonic mass ($f_\rmn{H} \approx 0.76$) and $m_\rmn{p}$ is the proton 
mass. From X-ray data of an ensemble of 45 clusters, \citet{1999ApJ...517..627M} derived $f_\rmn{B} = 0.075 h^{-3/2}$. 
Traditionally, the number density of dark matter haloes is described by the Press-Schechter formalism 
\citep{1974ApJ...187..425P}. The comoving Press-Schechter mass function can be written as
\begin{eqnarray}
  \label{eq:PS}
  n_\rmn{PS}(M,z) &=& \frac{\bar{\rho}}{\sqrt{2 \pi}D_{+}(z) M^2}\nonumber
  \left(1+\frac{n}{3}\right) \left(\frac{M}{M_*}\right)^{(n+3)/6}\\
  &\times& \exp\left[-\frac{1}{2 D_{+}^2(z)}
  \left(\frac{M}{M_*}\right)^{(n+3)/3}\right],
\end{eqnarray}
where $M_*$ and $\bar{\rho}$ are the nonlinear mass today and the mean background density at the present epoch, and 
$D_{+}(z)$ is the linear growth factor of density perturbations, normalised to unity today, $D_{+}(0) = 1$. 
$n \approx -1$ denotes the effective exponent of the dark matter power spectrum at the cluster scale. 
\citet{1999MNRAS.308..119S} recently proposed a significantly improved analytic derivation of the mass function while 
\citet{2001MNRAS.321..372J} measured the mass function of dark matter haloes in numerical simulations and found a fitting 
formula very close to Sheth \& Tormen's, however, being of slightly lower amplitude at high masses. Thus, the fitting formula 
found by Jenkins et al.\ was used in our study.

The total Compton-$y$ parameter per unit solid angle is given by
\begin{equation}
  \label{eq:Y}
  \mathcal{Y}_\Omega = \dang^2\int \dd^2 \theta \,y(\mathbf{\theta}) = 
  \frac{k_B T_\e}{m_\e c^2} \sigma_\rmn{T} N_\e,
\end{equation}
where $\dang$ is the angular-diameter distance to the cluster.  The mean background level of SZ fluctuations is given by
\begin{eqnarray}
  \label{eq:Ystatistics}
  \langle y_\rmn{bg} \rangle_\theta (z,M_0)  &=& 
  \int \dd z\left|\frac{\dd V}{\dd z}\right|(1+z)^3
  \int_{M_0} \dd M \,n_\rmn{PS}(M,z) \mathcal{Y}_\Omega(M,z)\nonumber\\
  &=& \int \dd M \int \dd V\, \mathcal{Y}_\Omega(M,z)\, 
  \frac{\dd^2 N(M,z)}{\dd M\, \dd V},
\end{eqnarray}
where $\dd V$ is the cosmic volume per unit redshift and unit solid angle, $n_\rmn{PS}(M,z)$ is the mass function of 
collapsed halos (\ref{eq:PS}), and $\mathcal{Y}_\Omega(M,z)$ is the integrated Compton-$y$ parameter per unit solid
angle from (\ref{eq:Y}) expressed in terms of halo mass $M$ and redshift $z$. Background fluctuations are due to Poisson 
fluctuations in the number of clusters per unit mass and volume if cluster correlations are neglected. The variance of 
the background fluctuations reads
\begin{equation}
  \label{eq:Y^2statistics}
  \langle y_\rmn{bg}^2 \rangle_\theta (z,M_0) =
  \int_{M_0} \dd M \int \dd V\, \left[\mathcal{Y}_\Omega(M,z)\right]^2\, 
  \frac{\dd^2 N(M,z)}{\dd M\, \dd V}.
\end{equation}

\begin{figure}
\resizebox{\hsize}{!}{\includegraphics{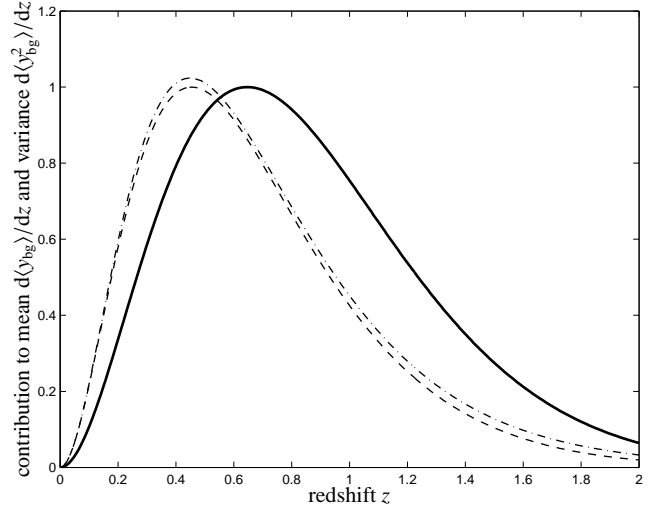}}
\caption{Comparison of the mean background level 
$\bra y_\rmn{bg} \ket_\theta (z,5\times 10^{13} M_{\sun}/h)$ (solid) and the variances
$\bra y_\rmn{bg}^2 \ket_\theta (z,5\times 10^{13} M_{\sun}/h)$ (dashed) and
$\bra y_\rmn{bg}^2 \ket_\theta (z,10^{13} M_{\sun}/h)$ (dash-dotted). The differential 
curves show qualitatively the smaller impact of low-mass and  high-redshift clusters on the variance compared to the 
mean background of SZ fluctuations.}
\label{fig:Y2}
\end{figure}

Fig.~\ref{fig:Y2} shows a qualitative comparison of the influence of the background of unresolved SZ-clusters on the 
mean background level $\bra y_\rmn{bg}\ket_\theta (z,M_0)$ and the variance $\bra y_\rmn{bg}^2 \ket_\theta (z,M_0)$. 
For studying this influence quantitatively, the ratio of mean background levels and variances is defined via:
\begin{eqnarray}
  \label{eq:ratio}
  r_\rmn{mean} & = &
  \frac{\langle y_\rmn{bg}\rangle_\theta (z_\mathrm{sim},M_\rmn{sim})}
  {\langle y_\rmn{bg}\rangle_\theta (z_\mathrm{max},M_\rmn{min})},\\  
  r_\rmn{var} & = &              
  \frac{\langle y_\rmn{bg}^2\rangle_\theta (z_\mathrm{sim},M_\rmn{sim})}
  {\langle y_\rmn{bg}^2\rangle_\theta (z_\mathrm{max},M_\rmn{min})},
\end{eqnarray}
where the numerator accounts for the resolved clusters in our simulation with $z<z_\rmn{sim}=1.5$ and 
$M>M_\rmn{sim}=5\times 10^{13} M_{\sun}/h$ while the denominator accounts for all collapsed halos contributing 
to the SZ-flux in our analytic estimate ($z<z_\rmn{max}=20$ and $M>M_\rmn{min}=1\times 10^{13} M_{\sun}/h$). 
Performing these integrals yields ratios of $r_\rmn{mean} = 40.6\%$ and $r_\rmn{var} = 93.3\%$ and thus confirms the 
qualitative picture of Fig.~\ref{fig:Y2}. Therefore, we conclude that we can safely neglect the effect of the background 
of unresolved SZ clusters on power spectrum statistics of our SZ all-sky map, especially when considering the mentioned 
uncertainties in $f_\rmn{B}$ and the ionisation fraction of electrons in low-mass halos.

\section{Results}\label{sect_results}
This section provides various characterisations of the SZ-cluster sample and properties of the resulting map. First, a 
visual impression of the SZ-maps is given in Sect.~\ref{sect_skyview}. Distribution of the angular sizes and of the 
integrated thermal and kinetic Comptonisations are presented in Sect.~\ref{sect_angsize} and in Sect.~\ref{sect_intwy}, 
respectively. The distribution of pixel amplitudes and a discussion of the sky-averaged thermal Comptonisation is given 
in Sect.~\ref{sect_pixamp}. The angular power-spectra in comparison to those obtained in high-resolution simulations 
performed by \citet{2002ApJ...579...16W} is shown in Sect.~\ref{sect_powerspec}. Finally, source counts in three 
relevant {\em Planck}-channels are given in Sect.~\ref{sect_sourcecount}.       

In order to quantify the deviations resulting in using template SZ-maps instead of relying solely on analytical 
profiles and idealised scaling relations, the distributions following from the respective approach are constrasted in  
Sect.~\ref{sect_angsize} (angular sizes), Sect.~\ref{sect_intwy} (integrated Comptonisations) and 
Sect.~\ref{sect_sourcecount} (source count at three selected \planck-frequencies).

\subsection{Sky views}\label{sect_skyview}
\begin{figure*}
\resizebox{\hsize}{!}{\includegraphics{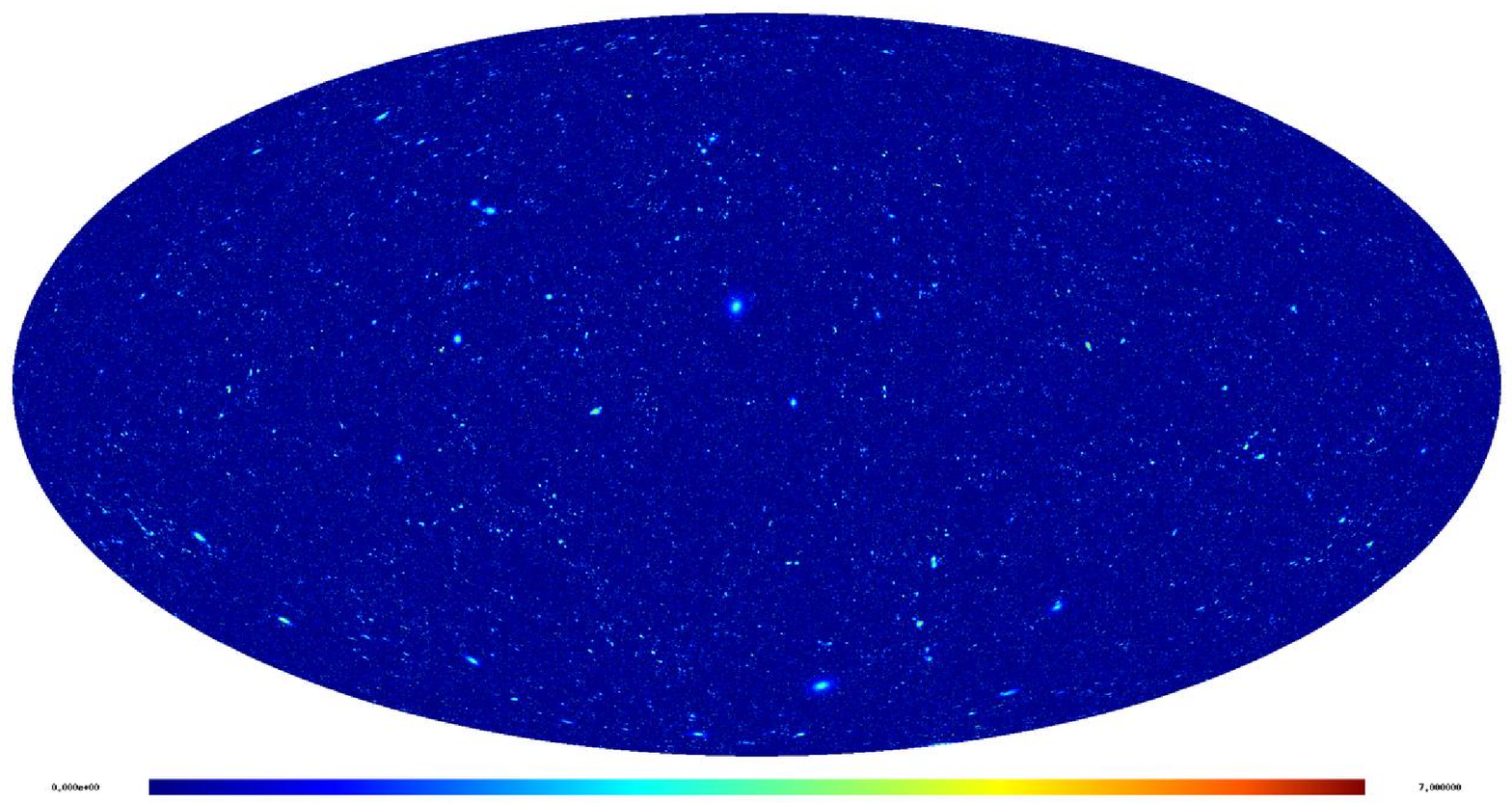}}
\caption{All-sky map of the thermal Comptonisation parameter $y$ in Mollweide projection. The shading is 
proportional to $\mathrm{arsinh}(10^6\times y)$.}
\label{fig_allsky_y}
\end{figure*}

\begin{figure*}
\resizebox{\hsize}{!}{\includegraphics{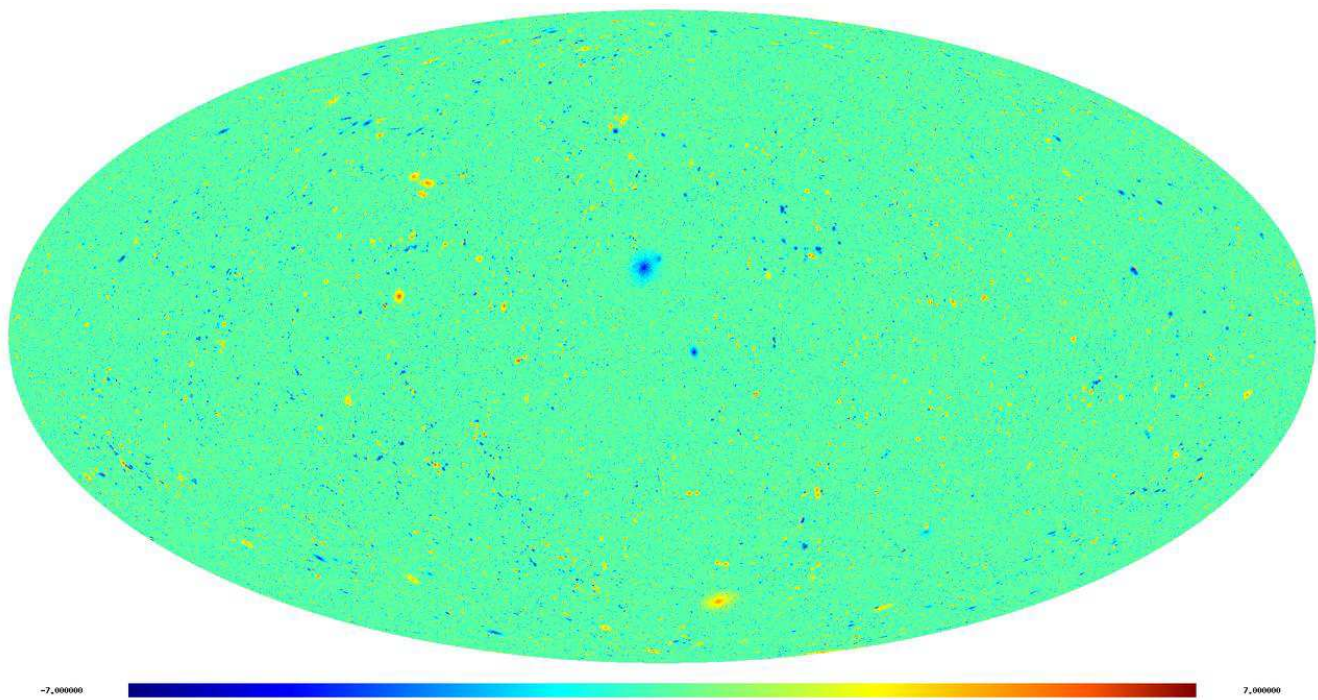}}
\caption{All-sky map of the kinetic Comptonisation parameter $w$ in Mollweide projection. The shading is 
proportional to $\mathrm{arsinh}(10^7\times w)$.}
\label{fig_allsky_w}
\end{figure*}

\begin{figure}
\resizebox{\hsize}{!}{\includegraphics{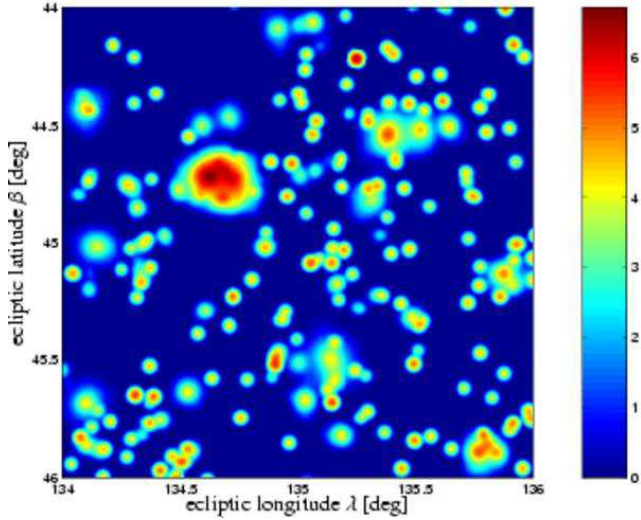}}
\caption{Detail of the thermal Comptonisation map: A $2\degr\times2\degr$ wide cut-out centered on the ecliptic 
coordinates $(\lambda,\beta) = (135\degr,45\degr)$ is shown. The smoothing imposed was a Gaussian kernel with 
$\Delta\theta = 2\farcm0$ (FWHM). The shading indicates the value of the thermal Comptonisation $y$ and is proportional to 
$\mathrm{arsinh}(10^6\times y)$. This map resulted from a projection on a Cartesian grid with mesh size $\sim14\arcsec$, i.e. 
no HEALPix pixelisation can be seen.}
\label{fig_cut_y}
\end{figure}

\begin{figure}
\resizebox{\hsize}{!}{\includegraphics{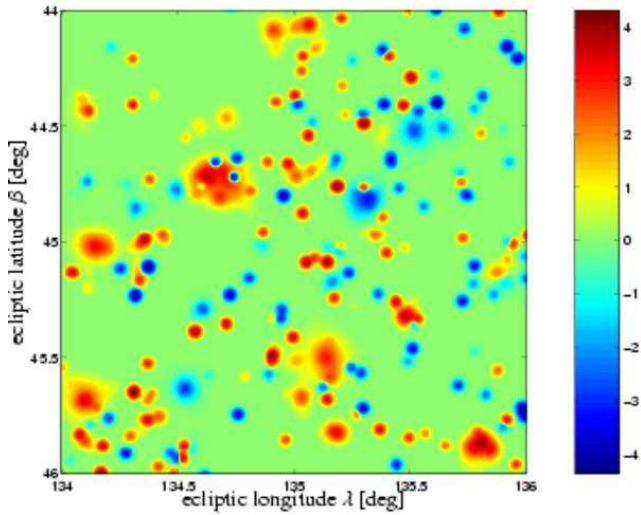}}
\caption{Detail of the kinetic Comptonisation map: A $2\degr\times2\degr$ wide cut-out centered on the same position as 
Fig.~\ref{fig_cut_y}, i.e. at the ecliptic coordinates $(\lambda,\beta) = (135\degr,45\degr)$ is shown. The smoothing 
imposed was a Gaussian kernel with $\Delta\theta = 2\farcm0$ (FWHM). The kinetic Comptonisation $w$ is indicated by the 
shading which is proportional to $\mathrm{arsinh}(10^6\times w)$.}
\label{fig_cut_w}
\end{figure}

In order to give a visual impression of the sky maps, all-sky views in Mollweide projection of the Compton-$y$ parameter 
(Fig.~\ref{fig_allsky_y}) as well as of the Compton-$w$ parameter (Fig.~\ref{fig_allsky_w}) are presented. Apart from 
those images, detailed maps of small regions of the SZ-sky are presented in Fig.~\ref{fig_cut_y} for the thermal and in 
Fig.~\ref{fig_cut_w} for the kinetic SZ-effects, respectively. These detailed maps display interesting features: 
Clearly, cluster substructure is visible in the maps, e.g. at position $(\lambda,\beta)\simeq(134\fdg60,45\fdg25)$. 

Secondly, massive clusters that generate a strong thermal signal, are rare, such that in drawing a peculiar velocity 
from a Gaussian distribution large values are less likely to be obtained. Consequently, these clusters commonly show only 
a weak kinetic signal, a nice example can be found at the position $(\lambda,\beta)\simeq(135\fdg25,44\fdg75)$. Closeby, the 
inverse example can be found at $(\lambda,\beta)\simeq(135\fdg40,44\fdg50)$, where a low-mass cluster shows only a weak 
thermal signal, but has sufficient optical depth and a high enough peculiar velocity to give rise to a strong kinetic signal. 
Finally, at $(\lambda,\beta)\simeq(135\fdg90,45\fdg75)$, there is an example of a merging cluster, with a dipolar variation of 
the subcluster velocities.

\begin{figure}
\resizebox{\hsize}{!}{\includegraphics{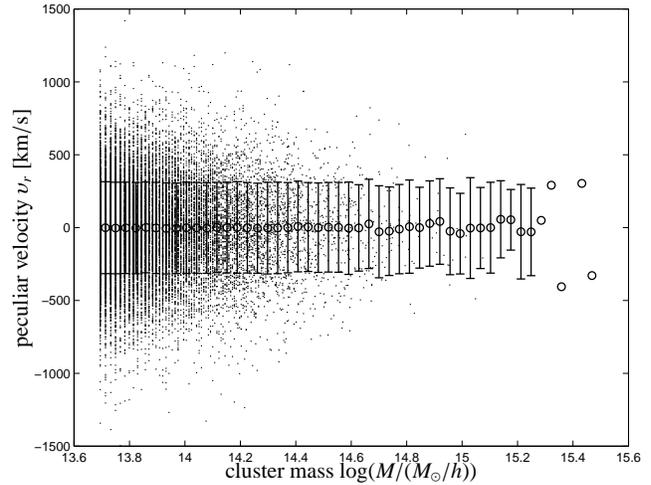}}
\caption{Mean value and variance of the (Gaussian) distribution of peculiar velocities $\upsilon_r$ as a function of cluster 
mass $M$. The parameters of the distribution do not depend on mass, i.e. the mean is consistent with zero and the standard 
deviation has values $\simeq320~\mathrm{km}/\mathrm{s}$ irrespective of mass. Standard deviations for the five bins 
corresponding to the largest cluster masses have been omitted due to poor statistics. The underlying data points represent 
1\% randomly selected entries of the Hubble-volume catalogue.}
\label{fig_pecvel}
\end{figure}

The occurence of high kinetic SZ-amplitudes is a subtle point: Cluster velocities follow a Gaussian distribution with 
mean consistent with zero, because the large scale structure is at rest in the comoving CMB-frame and with a standard 
deviation of $\sigma_\upsilon = 312.8\pm0.2~\mathrm{km}/\mathrm{s}$. This value has been measured for clusters in 
the Hubble-volume catalogue and is noticably smaller compared to $\sigma_\upsilon = 400~\mathrm{km}/\mathrm{s}$ 
proposed by \citet{1997A&A...325....9A}. As Fig.~\ref{fig_pecvel} illustrates, the velocity-distribution does {\em not} depend 
on the cluster mass, because on the scales of typical cluster separation, linear structure formation is responsible for  
accelerating the clusters to their peculiar velocity. Massive clusters are rare and thus a high peculiar velocity is 
seldomly drawn from the underlying Gaussian distribution. Despite the seemingly large separation, it would be 
incorrect to draw the velocities independently from a Gaussian distribution. Instead, the kinetic SZ-map ensures the 
consistency that the density and velocity fields have grown from the initial Gaussian random field by linear structure 
formation and have the correct relative phases.

Similar to the clustering on large angular scales that the thermal SZ-map shows due to the formation of superclusters, 
the kinetic SZ-map is expected to exhibit clustering on the same angular scales. This is because in the formation of 
superclusters, the velocity vectors of infalling clusters point at the dynamical centre and are thus correlated despite 
the large separation.

\subsection{Distribution of angular sizes}\label{sect_angsize}
The distribution of cluster sizes is an important characteristic of the sky maps. For the derivation of core sizes, two 
different paths have been pursued in order to contrast the ideal case, in which cluster sizes follow from the well-known 
virial relations to the simulated and realistic case, in which the sizes are measured on the template maps themselves.
First, the cluster sizes are measured on the data by fitting a King-profile \citep{1978A&A....70..677C} to the 
thermal and kinetic template maps:
\begin{eqnarray}
y\left(\bmath{r}\right) & = & y_0\cdot\left[1 + \left(\frac{\left|\bmath{r}\right|}{r_c^{(y)}}\right)^2\right]^{-1} \\
w\left(\bmath{r}\right) & = & w_0\cdot\left[1 + \left(\frac{\left|\bmath{r}\right|}{r_c^{(w)}}\right)^2\right]^{-1}
\mbox{,}
\end{eqnarray}
yielding the core radii $r_c^{(y)}$ for the thermal and $r_c^{(w)}$ for the kinetic map, respectively. The 
$\beta$-fits have been centered on the pixel with the highest amplitude, and as free parameters only the central 
amplitudes $y_0$ and $w_0$ were used apart from the core radii. The resulting radii have been averaged over all three 
projections of the cluster. Together with the comoving distance of the cluster as given by the Hubble-volume catalogue and the 
scaling factor required to match the size (compare Sect.~\ref{map_scaling}), the core radii have been converted into 
angular diameters.

Secondly, an angular extent has been derived from the virial radius. Template data suggests the relation
\begin{equation}
r_c \simeq 0.12\, r_\mathrm{vir}
\end{equation}
rather than the value of $r_c\simeq 0.07\, r_\mathrm{vir}$ advocated by \citet{2000MNRAS.315..689L} and 
\citet{2001MNRAS.325..835K}. In analogy, the angular diameter was then determined with the cluster distance given by 
the Hubble-volume catalogue.
                                 
\begin{figure}                   
\resizebox{\hsize}{!}{\includegraphics{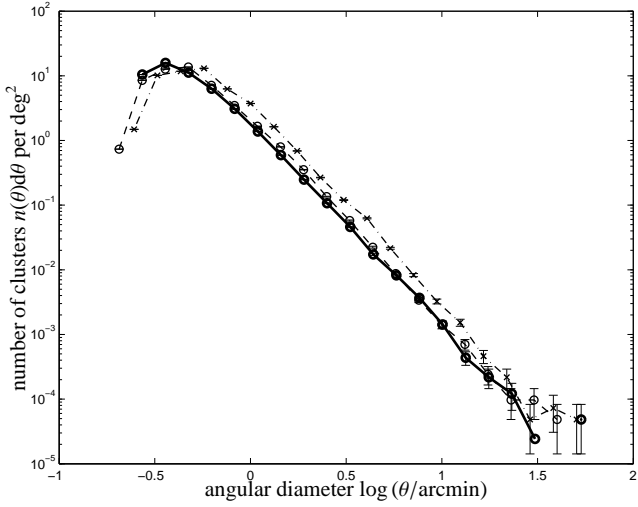}}
\caption{The number of clusters $n(\theta)\dd\theta$ per square degree for given angular diameter $\theta$ is shown 
without taking beam convolution into account, for thermal clusters (circles, dashed line) and kinetic clusters (crosses, 
dash-dotted line) as following from $\beta$-model fits to template data. For comparison, the distribution of angular 
diameters obtained via the virial theorem (solid line) is also plotted.}
\label{fig_size_distr}
\end{figure}

In Fig.~\ref{fig_size_distr}, the size distributions for the thermal as well as for the kinetic clusters are given. 
Clearly, most clusters have angular diameters small compared to \plancks beam, and would appear as point sources. 
Here, it should be emphasised, that the HEALPix tesselation with the chosen $N_\mathrm{side}$-parameter does not 
resolve structures smaller than $1\farcm71$. In the process of smoothing the clusters imposed prior to projection 
(compare Sect.~\ref{map_project}), clusters with diameters smaller than $2\farcm0$ have been replaced by 2-dimensional 
Gaussians with $\sigma = 2\farcm0$. Their normalisation corresponds to the integrated Comptonisations 
$\mathcal{Y}$ and $\mathcal{W}$ measured on the template maps. The smoothing is an absolute necessity because 
otherwise the HEALPix map would need $\sim10^4$ times as many pixels for supporting the most distant and hence smallest 
clusters in the Hubble sample and hence $\sim10^4$ times the storage space. A futher point to notice is the remarkably 
good agreement between diameters derived from the various prescriptions.

\subsection{Distribution of the integrated thermal and kinetic Comptonisation}\label{sect_intwy}
The signal strength of a cluster in an SZ observation is not given by the line-of-sight Comptonisation, but rather the 
Comptonisation integrated over the solid angle subtended by the cluster. These quantities are refered to as the 
integrated thermal Comptonisation $\mathcal{Y}$ and kinetic Comptonisation $\mathcal{W}$ and are defined as:
\begin{equation}
\mathcal{Y} = \int\dd\Omega\, y(\bmath{\theta})\quad\mbox{and}\quad 
\mathcal{W} = \int\dd\Omega\, w(\bmath{\theta})\mbox{.}
\end{equation}

For a simple model of the integrated Comptonisations as functions of cluster mass $M$, distance $z$ and peculiar 
velocity $\upsilon_r$ it is assumed that the SZ-flux originates from inside a sphere of radius $r_\mathrm{vir}$, 
that the baryon fraction is equal to its universal value $f_B = \Omega_B/\Omega_M$, that the ICM is completely ionised 
and has a uniform temperature predicted by the spherical collapse model laid down in eqn.~(\ref{eqn_virial_temp}). In 
this model, the actual distribution of electrons inside the virial sphere is of no importance. Then, the integrated 
Comptonisations are approximated by:
\begin{eqnarray}
\frac{\mathcal{Y}_\mathrm{vir}}{\mathrm{arcmin}^2}& = & 1.98\,\frac{f_B}{h}
\left(\frac{M_\mathrm{vir}}{M_\star}\right)^\frac{5}{3}
\left(\frac{d_A}{d_\star}\right)^{-2}\!(1+z)
\left(\frac{\Omega_0}{\Omega}\right)^\frac{1}{3}
\left(\frac{\Delta_\rmn{c}}{178}\right)^\frac{1}{3}\! ,\label{eqn_y_scaling}\\
\frac{\mathcal{W}_\mathrm{vir}}{\mathrm{arcmin}^2} & = & 0.29\,\frac{f_B}{h}
\left(\frac{M_\mathrm{vir}}{M_\star}\right)
\left(\frac{d_A}{d_\star}\right)^{-2}
\left(\frac{\upsilon_r}{\upsilon_\star}\right)\mbox{,}
\label{eqn_w_scaling}
\end{eqnarray}
respectively. The reference values have been chosen to be $M_\star = 10^{15} M_{\sun}/h$, $d_\star = 100\mbox{ Mpc}/h$ 
and $\upsilon_\star = 1000\mbox{ km}/\mbox{s}$. $d_A$ is the angular diameter distance to the cluster. $\Omega=\Omega(z)$ 
denotes the mass density at redshift $z$ and $\Delta_\mathrm{c}=\Delta_\mathrm{c}(z)$ the overdensity of a virialised 
sphere, an approximate description is given by eqn.~(\ref{eqn_overdensity}). For typical values for mass, distance and 
velocity, the thermal and kinetic SZ-effects differ by approximately one order of magnitude. The baryon fraction is set 
to the universal value $f_B = \Omega_B/\Omega_M = 0.133$ for the remainder of this paper.

\begin{figure}
\resizebox{\hsize}{!}{\includegraphics{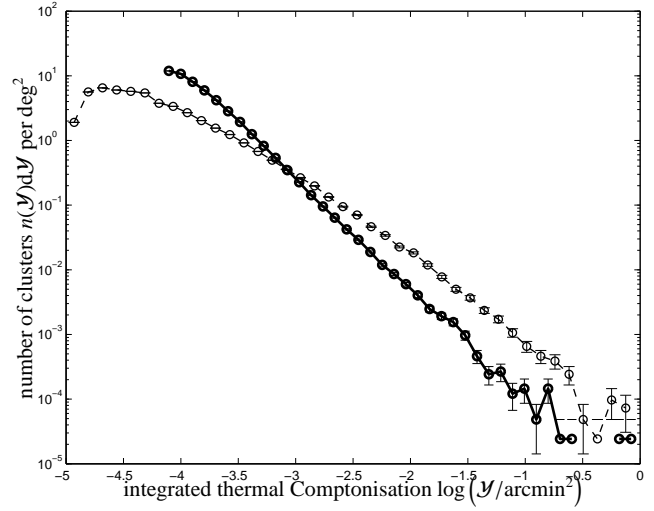}}
\caption{The number of clusters $n(\mathcal{Y})\dd\mathcal{Y}$ per square degree for given integrated thermal 
Comptonisation $\mathcal{Y}$ derived from template data (dashed line) in comparison to the analogous quantitiy 
based on virial estimates (solid line).}
\label{fig_ydistr}
\end{figure}

\begin{figure}
\resizebox{\hsize}{!}{\includegraphics{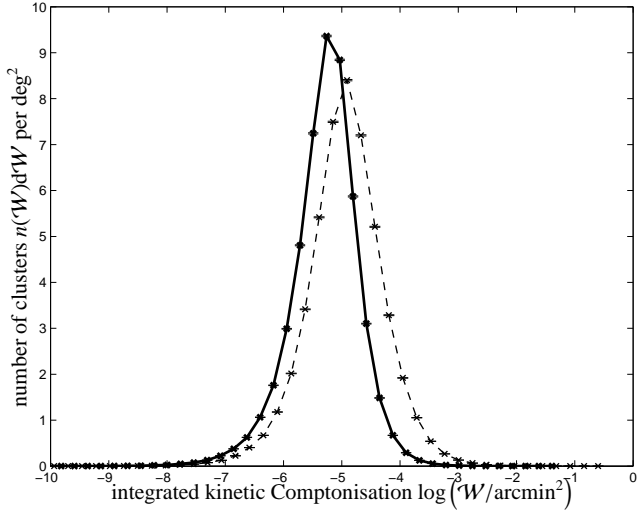}}
\caption{The number of clusters $n(\mathcal{W})\dd\mathcal{W}$ per square degree for given integrated kinetic 
Comptonisation $\mathcal{W}$ derived from template data (dashed line) in comparison to the analogous quantitiy 
based on virial estimates (solid line). Here, the scaling of the vertical axis is linear, in contrast to 
Fig.~\ref{fig_wdistr}, such that the underlying Gaussian distribution of peculiar velocities becomes apparent.}
\label{fig_wdistr}
\end{figure}

Distributions of the integrated thermal and kinetic Comptonisations are shown in Figs.~\ref{fig_ydistr} and 
\ref{fig_wdistr}, respectively. The distributions have been derived from actual scaled template data in comparison to 
the values obtained from~(\ref{eqn_y_scaling}) and (\ref{eqn_w_scaling}).

Fig.~\ref{fig_ydistr} shows the number of clusters per $\mathrm{deg}^2$ with integrated thermal Comptonisation 
$\mathcal{Y}$. It can be seen that the approach relying on the virial theorem underestimates the number of cluster by a 
factor 2-3 for large integrated Comptonisations. Alternatively, one could state that the distributions are separated at 
high Comptonisations by slightly less than 0.5~dex. The reason for the significantly larger integrated Comptonisations 
determined from template data is due to the fact that the template clusters were matched to the catalogue entries given by the 
Hubble-volume simulation by their virial masses. In irregular clusters, there is a significant fraction of the gas located 
outside the virial sphere and thus the integrated Comptonisation is systematically underestimated when applying a spherical 
overdensity code to simulation data, as previously examined in Fig.~\ref{fig_qmqt_distribution}. In scaling the template 
clusters up to the masses required by the Hubble-volume catalogue, this difference is amplified because in the sparsely sampled 
region of the $M$-$z$-diagram (compare Figs.~\ref{figure_mzplane} and \ref{fig_qmqt_distribution}) clusters have on average to 
be scaled to higher masses, which explains the offset in the distributions. 
A second effect is the evolution of ICM temperature. Compared to the temperature model eqn.~(\ref{eqn_virial_temp}) based 
on spherical collapse theory, the plasma temperatures are smaller by approximately 25\%, i.e. the mean SPH-temperature 
of the particles inside the virial sphere is smaller than expected from spherical collapse theory and 
reflects the departure from isothermality: The template clusters do show a temperature profile that declines towards the 
outskirts of the clusters, which decreases the integrated Comptonisation relative to the values derived by means 
of the virial theorem. 

Furthermore, the dependence of electron temperature on cluster mass is noticably weaker than the $M^{2/3}$-scaling: The 
cluster number weighted average for the exponent $\alpha$ in the scaling $T\propto M^\alpha$ relating temperature to 
mass was found to be $\bra\alpha\ket=0.624$, and at the redshifts around unity, where  most of the clusters reside, 
values as small as $\alpha=0.605$ were derived. Using this scaling, the Compton-$y$ parameter and hence the integrated 
thermal Comptonisation $\mathcal{Y}$ shows a significantly shallower distribution compared to the distribution relying 
on simple scaling arguments.

The same argument applies to the kinetic Comptonisation $\mathcal{W}$, as depicted in Fig.~\ref{fig_wdistr}: Here a 
shifting of the values to smaller kinetic Comptonisations is observed when comparing estimates following from the virial 
theorem to actual simulation data. The shift of the peak of the distribution amounts to about one third dex, 
as explained above for the thermal Comptonisations. Keeping the $M^\frac{5}{3}$-scaling of the thermal SZ-effect in 
mind, the shift in the $\mathcal{W}$-distribution is then consistent with the shift of the $\mathcal{Y}$-distribution.

\subsection{Distribution of Comptonisation per pixel}\label{sect_pixamp}
\begin{figure}
\resizebox{\hsize}{!}{\includegraphics{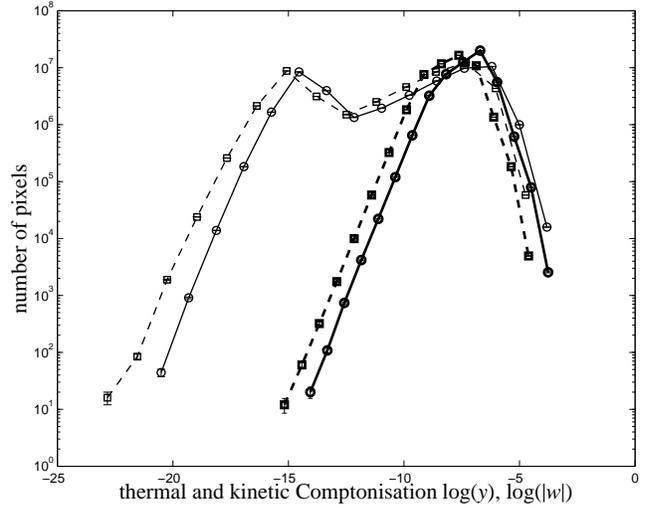}}
\caption{Distribution of pixel amplitudes of the thermal Comptonisation parameter $y$ (circles, solid) and of the 
absolute value of the kinetic Comptonisation $\left|w\right|$ (squares, dashed). Zero values have been deliberately 
excluded. The figure illustrates the occurence of clusters comparable to the HEALPix pixel scale: The thin set of 
lines shows an additional peak at small Comptonisations, that vanish after convolution with a beam of $\Delta\theta = 
5\farcm0$ (FWHM), i.e. comparable to \planck, as shown by the thick lines.}
\label{fig_mean_y}
\end{figure}

Fig.~\ref{fig_mean_y} shows the distribution of the pixel amplitudes of the thermal SZ-map as well as of their absolute values 
in the kinetic SZ-map. Clearly, the kinetic and thermal SZ-effects are separated by approximately one order of magnitude. 

The distribution of pixel amplitudes is very broad, encompassing the largest line-of-sight Comptonisations of 
$y\simeq1.5\times10^{-4}$ and $\left|w\right|\simeq1.6\times10^{-5}$ down to very low signals below $\log(y)\simeq-20$. 
The distribution is bimodal, which is a pixelisation artefact and which is caused by the replacement of faint and small 
clusters with a very narrow Gaussian, the extent of which is slightly above the pixel scale, once the cluster is 
smaller than $\simeq2\farcm0$ in diameter. These clusters are more concentrated than the King-profiles of resolved 
clusters. There is a caveat when applying an expansion into spherical harmonics $Y^\ell_m(\bmath{\theta})$ to the SZ-maps: 
The smallest clusters are only a few pixels in diameter. Working with the HEALPix tesselation, reliable expansion 
coefficients can only be obtained up to multipole moments of order $\ell\simeq2\times N_\mathrm{side}$, i.e. up to 
$\ell\simeq 4096$ in our case, which corresponds to angular scales of $2\farcm64$. The pixel scale is $1\farcm71$ for 
this choice of the $N_\mathrm{side}$-parameter. Consequently, the small clusters will not be contained in an expansion 
into spherical harmonics, as shown by the set of thick lines in Fig.~\ref{fig_mean_y}: Here, the map has been 
decomposed in $a_{\ell m}$-coefficients (compare eqn.~(\ref{eqn_ylm_decomp})), multiplied with the $a_{\ell 
0}$-coefficients of a Gaussian beam of $\Delta\theta = 5\farcm0$ (FWHM) and synthesised again. Then, the resulting 
smoothed map does not contain small clusters, because the decomposition into spherical harmonics has not been able to 
resolve structures that extend over only a few pixels.

The mean value of the thermal Comptonisation $y$ has been determined to be $\bra y\ket=3.01\times10^{-7}$ and the 
pixel-to-pixel variance is $\sigma_y=\sqrt{\bra y^2\ket-\bra y\ket ^2}=1.85\times10^{-6}$. In analogy, the value 
$\bra w\ket=6.28\times10^{-9}$ has been derived for the kinetic map, with variance 
$\sigma_w=\sqrt{\bra w^2\ket-\bra w\ket ^2}=3.78\times10^{-7}$, i.e. the mean kinetic Comptonisation is consistent 
with zero, due to the peculiar velocities following a Gaussian distribution with zero mean. The mean value of the moduli of 
the pixel amplitudes of the kinetic map is $\bra\left|w\right|\ket=7.65\times10^{-8}$.

The value for the mean Comptonisation $\bra y\ket$ measured on the map should account for roughly 40\% of the mean 
thermal Comptonisation as derived in Sect.~\ref{sect_szbg}, due to the lower mass threshold inherent to the simulation. 
Keeping in mind the absence of any diffuse component of the thermal Comptonisation, the value derived here is compatible 
with the value of $\simeq 10^{-6}$ given by \citet{2000PhRvD..61l3001R} and \citet{1993ApJ...416..399S}, but falls short 
of the value derived by \citet{2002ApJ...579...16W} by a factor of less than two. \citet{2004MNRAS.347L..67M} performed 
a cross-correlation of WMAP-data with clusters from the APM survey and found the mean Comptonisation to be significantly 
larger and to be in accordance with \citet{1995ApJ...442....1P}, but in contradiction with expectations from CDM models.

\subsection{Angular power spectra of the thermal and kinetic SZ-effects}\label{sect_powerspec}
\begin{figure}
\resizebox{\hsize}{!}{\includegraphics{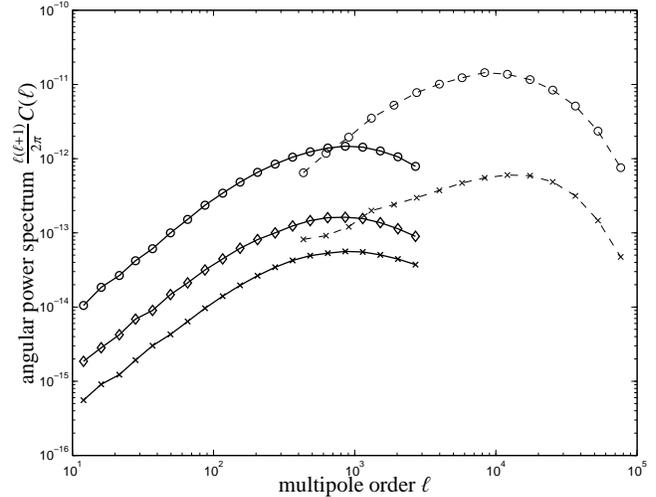}}
\caption{Angular power spectra of the thermal and kinetic SZ-effect: $C_{yy}(\ell)$ (circles, solid line), 
$C_{ww}(\ell)$ (crosses, solid line) and the cross power spectrum $C_{yw^\prime}(\ell)$ (diamonds, solid line) are shown 
in comparison to the power spectra of the thermal SZ-effect (circles, dashed line) and the kinetic SZ-effect (crosses, 
dashed line) obtained by \citet{2002ApJ...579...16W} at smaller scales, i.e. at higher multipole order $\ell$.}
\label{fig_power_spectra}
\end{figure}

In this section, the angular power spectra are given for the all-sky maps. They follow from a decomposition of the 
spherical data set into spherical harmonics $Y_\ell^m(\bmath{\theta})$:
\begin{eqnarray}
y_{\ell m} & = & \int_{4\pi}\dd\Omega\, y(\bmath{\theta})\, Y_\ell^m(\bmath{\theta})^*\mbox{,} \\
w_{\ell m} & = & \int_{4\pi}\dd\Omega\, w(\bmath{\theta})\, Y_\ell^m(\bmath{\theta})^*\mbox{, and } \\
w^\prime_{\ell m} & = & \int_{4\pi}\dd\Omega\, \left|w(\bmath{\theta})\right|\, 
Y_\ell^m(\bmath{\theta})^*\mbox{,}
\label{eqn_ylm_decomp}
\end{eqnarray}
respectively. The spherical harmonical transform $w^\prime_{\ell m}$ has been determined from the absolute values of the 
kinetic map amplitudes. The reason for doing so is the vanishing expectation value of the peculiar velocities in the 
comoving frame such that for a given cluster in the thermal SZ-map, both signs of the kinetic SZ-effect are equally 
likely to occur and the cross-power averages out to zero.
The angular power spectra and the cross power spectrum are defined via:
\begin{eqnarray}
C_{yy}(\ell) & = & \frac{1}{2\ell+1}\sum_{m=-\ell}^{+\ell} y_{\ell m}\, y_{\ell m}^*\mbox{,} 
\label{eqn_yspec} \\
C_{ww}(\ell) & = & \frac{1}{2\ell+1}\sum_{m=-\ell}^{+\ell} w_{\ell m}\, w_{\ell m}^*\mbox{,} 
\label{eqn_wspec} \\
C_{yw^\prime}(\ell) & = & \frac{1}{2\ell+1}\sum_{m=-\ell}^{+\ell} w^\prime_{\ell m}\, y^*_{\ell m}
\label{eqn_cross}\mbox{,}
\end{eqnarray}
with the asterisk denoting complex conjugation. The resulting power spectra are given in Fig.~\ref{fig_power_spectra} 
in comparison to the power-spectra derived by \citet{2002ApJ...579...16W} in simulations covering smaller angular 
scales. The curves match well, and the remaining discrepancies may be explained by the fact that in the maps presented 
here, power is missing on small scales due to the low-mass cutoff, whereas the simulation by White is missing power 
on large scales due to the smallness of their simulation box. The bending-over of the spectra derived from our SZ-maps is also 
due to the fact that the expansion in spherical harmonics cannot be computed for angular scales approaching the pixel scale and 
thus does not include very small clusters of sizes comparable to the pixel size, as already discussed in 
Sect.~\ref{sect_pixamp}.

If clusters were randomly positioned point-sources on the sky, the number of clusters per solid angle element would be a 
Poisson-process and the resulting power spectrum should be flat, i.e. $C(\ell)\propto N$ (N ist the number of sources), as 
shown by \citet{2001PhRvD..63f3001S}. Contrarily, the brightness distribution of clusters assigns additional weight to the 
large angular scales and giving rise to a significant deviation in the slope of the power spectra 
$C(\ell)\propto\ell^{-\gamma}$ as a function of $\ell$: The measured slope is $2+\gamma=1.53\pm0.07$ for the thermal and 
$2+\gamma=1.45\pm0.07$ for the kinetic SZ-effect, which reflects the deviation from pure Poissonianity. In the fitting, the 
values for $\ell$ have been restricted to $1\leq\ell\leq100$ and the errors derived correspond to the 95\% confidence 
intervals.

Furthermore, Fig.~\ref{fig_power_spectra} shows the cross-correlation between the thermal SZ-map and the absolute value 
of the kinetic SZ-map. As expected, the amplitude of the cross-power spectrum is at an intermediate level compared to 
autocorrelations of the thermal and kinetic SZ-maps.

\subsection{Source counts at Planck frequencies}\label{sect_sourcecount}
As the last point in this analysis, we address the SZ source counts, i.e.~the number $N$ of SZ-clusters giving rise to flux 
changes exceeding a certain flux threshold $S_\mathrm{min}$. The SZ flux modulation as a function of frequency is given 
by:
\begin{eqnarray}
S(x) & = & S_0\int\dd\Omega\,\left[y\, g(x) - \beta\tau\, h(x)\right] \\ 
     & = & S_0\left[\mathcal{Y} g(x)- \mathcal{W} h(x)\right] = S_\mathcal{Y}(x) - S_\mathcal{W}(x)\mbox{,}
\label{eqn_szflux}
\end{eqnarray}
where $S_0 = 22.9\mbox{ Jy}/\mbox{arcmin}^2$ is the flux density of the CMB and $\mathcal{Y}$ and $\mathcal{W}$ denote 
the integrated thermal and kinetic Comptonisations. The functions $g(x)$ and $h(x)$ are the flux modulations caused by 
the thermal and kinetic SZ-effects for non-relativistic electron velocities:
\begin{eqnarray}
g(x) & = & \frac{x^4\exp(x)}{(\exp(x)-1)^2}\left[x\frac{\exp(x)+1}{\exp(x)-1}-4\right]\mbox{,} \\
h(x) & = & \frac{x^4\exp(x)}{(\exp(x)-1)^2}\mbox{.}
\end{eqnarray}
Here, $x$ again denotes the dimensionless frequency $x = h\nu / (k_B T_\mathrm{CMB})$. The averaged flux 
$\bra S\ket_{\nu_0}$ at the fiducial frequency $\nu_0$ is obtained by weighted summation with the frequency response 
window function $R_{\nu_0}(\nu)$ and can readily be converted to antenna temperature $T_A$ by means of 
eqn.~(\ref{eqn_antenna_temp}):
\begin{equation}
\bra S\ket_{\nu_0} = 
\frac{\int\dd\nu\,S(\nu)\, R_{\nu_0}(\nu)}{\int\dd\nu\,R_{\nu_0}(\nu)} = 
2\frac{\nu_0^2}{c^2}\, k_B\,T_A\mbox{.}
\label{eqn_antenna_temp}
\end{equation}

The main characteristics of \plancks receivers and the conversion factors from 1~$\mathrm{arcmin}^2$ of thermal or 
kinetic Comptonisation to fluxes in Jansky and changes in antenna temperature measured in \nK~is given by 
Table~\ref{table_planck_channel}. For the derivation of the values a top-hat shaped frequency response function 
$R_{\nu_0}(\nu)$ has been assumed:
\begin{equation}
R_{\nu_0}(\nu) =
\left\{
\begin{array}{l@{,\:}l}
1  & \nu\in\left[\nu_0-\Delta\nu,\nu_0+\Delta\nu\right] \\
0  & \nu\notin\left[\nu_0-\Delta\nu,\nu_0+\Delta\nu\right] 
\end{array}
\right.\mbox{.}
\label{eqn_freq_window}
\end{equation}

\begin{table*}
\vspace{-0.1cm}
\begin{center}
\begin{tabular}{lrrrrrrrrr}
\hline\hline
\vphantom{\Large A}%
{\em Planck} channel	& 1 & 2 & 3 & 4 & 5 & 6 & 7 & 8 & 9 \\
\hline
\vphantom{\Large A}%
centre frequency $\nu_0$				
& 30~GHz   & 44~GHz   & 70~GHz   & 100~GHz & 143~GHz & 217~GHz & 353~GHz & 545~GHz & 857~GHz 		\\
frequency window $\Delta\nu$			
& 3.0~GHz & 4.4~GHz & 7.0~GHz & 16.7~GHz & 23.8~GHz & 36.2~GHz & 58.8~GHz & 90.7~GHz & 142.8~GHz	\\
thermal SZ flux $\bra S_\mathcal{Y}\ket$	
& -12.2~Jy & -24.8~Jy & -53.6~Jy & -82.1~Jy & -88.8~Jy &  -0.7~Jy & 146.0~Jy & 76.8~Jy  & 5.4~Jy	\\
kinetic SZ flux $\bra S_\mathcal{W}\ket$
& 6.2~Jy   & 13.1~Jy  & 30.6~Jy  &  55.0~Jy &  86.9~Jy & 110.0~Jy &  69.1~Jy & 15.0~Jy  & 0.5~Jy	\\
antenna temperature $\Delta T_\mathcal{Y}$ 	
& -440~\nK & -417~\nK & -356~\nK & -267~\nK & -141~\nK &  -0.5~\nK&   38~\nK &  8.4~\nK & 0.2~\nK	\\
antenna temperature $\Delta T_\mathcal{W}$ 	
& 226~\nK  & 220~\nK  & 204~\nK  &  179~\nK &  138~\nK &    76~\nK&   18~\nK &  1.6~\nK & 0.02~\nK	\\
\hline
\end{tabular}
\end{center}
\caption{Characteristics of \plancks LFI- and HFI- receivers: centre frequency $\nu$, frequency window $\Delta\nu$ (as 
defined in eqn.~(\ref{eqn_freq_window})), fluxes $\bra S_\mathcal{Y}\ket$ and $\bra S_\mathcal{W}\ket$ (see 
eqn.~(\ref{eqn_szflux})) generated by the respective Comptonisation of $\mathcal{Y} = \mathcal{W} = 1~\mathrm{arcmin}^2$ and 
the corresponding changes in antenna temperature $\Delta T_\mathcal{Y}$ and $\Delta T_\mathcal{W}$. Due to \plancks symmetric 
frequency response window, the thermal SZ-effect does not vanish entirely at $\nu=217$~GHz.}
\label{table_planck_channel}
\end{table*}

\begin{figure}
\resizebox{\hsize}{!}{\includegraphics{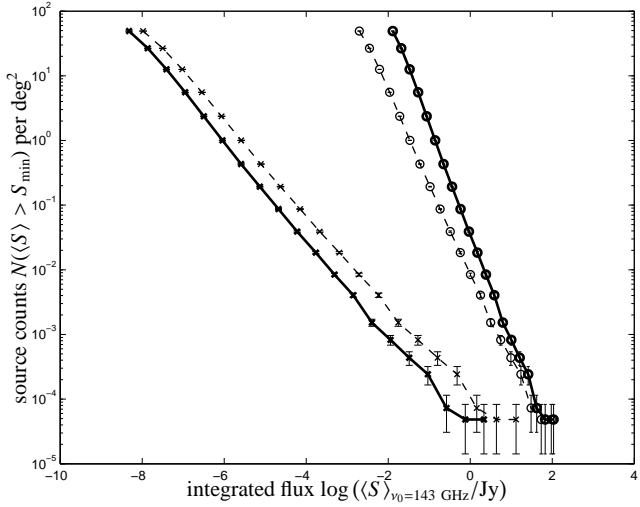}}
\caption{Source counts $N(\bra S\ket>S_\mathrm{min})$ for thermal (circles) and kinetic clusters (crosses) for 
\plancks $\nu_0 = 143$~GHz channel and for fluxed measured on the scaled template clusters (dashed line) in 
comparison to virial fluxes (solid line).}
\label{fig_sc_143}
\end{figure}

\begin{figure}
\resizebox{\hsize}{!}{\includegraphics{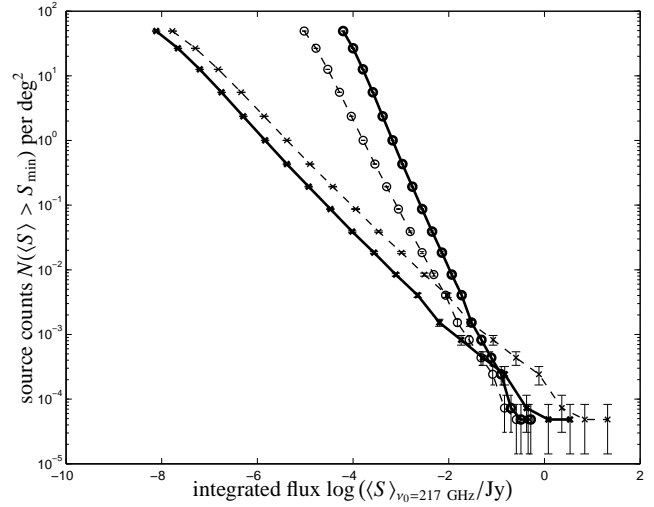}}
\caption{Source counts $N(\bra S\ket>S_\mathrm{min})$ for thermal (circles) and kinetic clusters (crosses) for 
\plancks $\nu_0 = 217$~GHz channel, the dashed and solid lines contrast the fluxes measured on the template data and 
those following from virial scaling relations, respectively. For the given frequency response function $R_{\nu_0}(\nu)$, 
the thermal SZ-effect does not vanish entirely at $\nu_0=217$~GHz.}
\label{fig_sc_217}
\end{figure}

\begin{figure}
\resizebox{\hsize}{!}{\includegraphics{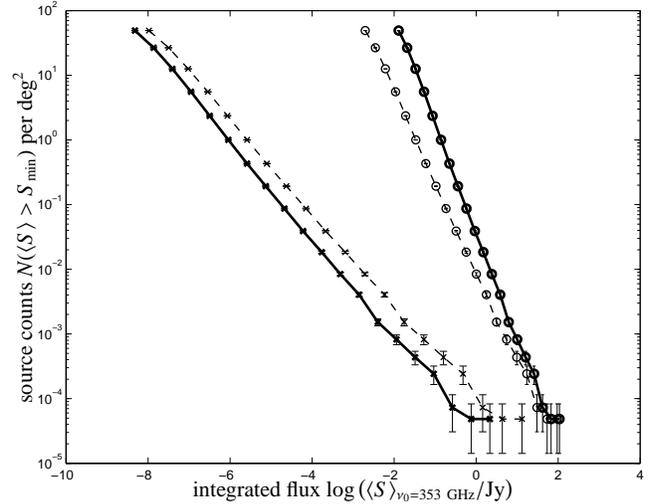}}
\caption{Source counts $N(\bra S\ket>S_\mathrm{min})$ for thermal (circles) and kinetic clusters (crosses) for 
\plancks $\nu_0 = 353$~GHz channel, again for fluxes derived from template data (dashed line) in comparison to 
fluxes following from virial scaling relations (solid line).}
\label{fig_sc_353}
\end{figure}

Figures~\ref{fig_sc_143}, \ref{fig_sc_217} and \ref{fig_sc_353} show the source counts stated in number of clusters per
$\mathrm{deg}^2$ as a function of averaged flux $\bra S\ket_{\nu_0}$ for \plancks $\nu_0=143$~GHz-, $\nu_0=217$~GHz- 
and $\nu_0=353$~GHz-channels, respectively. 

The source counts $N(S)$ are well approximated by power laws of the form:
\begin{equation}
N(\bra S\ket>S_\mathrm{min}) = N_0\, S^\alpha\mbox{.}
\label{eqn_power_source_count}
\end{equation}
Values for the normalisations $N_0$ and the slopes $\alpha$ have been obtained by fits to the source counts 
for the three relevant {\em Planck}-frequencies and are stated in Table~\ref{table_source_count_fits}. In the fits, the 
four rightmost bins have been excluded because of poor statistics. The parameters of the power law has been derived for 
the fluxes following from the idealised case based on the virial theorem and compared to fluxes determined from template 
cluster data. 

\begin{table*}
\vspace{-0.1cm}
\begin{center}
\begin{tabular}{lrrr}
\hline\hline
\vphantom{\Large A}%
{\em Planck} channel 			& $\nu_0=143$~GHz & $\nu_0=217$~GHz & $\nu_0=353$~GHz \\
\hline
\vphantom{\Large A}%
thermal SZ-effect, virial estimate	& $\log N_0=-1.78\pm0.02$ & $\log N_0=-5.31\pm0.02$ & $\log N_0=-1.42\pm0.22 $\\
					& $\alpha=-1.67\pm0.03$ & $\alpha=-1.67\pm0.07$ & $\alpha=-1.66\pm0.02 $\\
kinetic SZ-effect, virial estimate	& $\log N_0=-4.49\pm0.01$ & $\log N_0=-4.42\pm0.01$ & $\log N_0=-4.57\pm0.01 $\\
					& $\alpha=-0.76\pm0.06$ & $\alpha=-0.76\pm0.05$ & $\alpha=-0.76\pm0.06 $\\
thermal SZ-effect, simulation		& $\log N_0=-2.36\pm0.02$ & $\log N_0=-5.31\pm0.02$ & $\log N_0=-2.06\pm0.02 $\\
					& $\alpha=-1.40\pm0.03$ & $\alpha=-1.40\pm0.06$ & $\alpha=-1.40\pm0.03 $\\
kinetic SZ-effect, simulation		& $\log N_0=-3.95\pm0.01$ & $\log N_0=-3.88\pm0.01$ & $\log N_0=-4.02\pm0.01 $\\
					& $\alpha=-0.72\pm0.04$ & $\alpha=-0.72\pm0.05$ & $\alpha=-0.72\pm0.05 $\\
\hline
\end{tabular}
\end{center}
\caption{Values obtained from fits of a power law of the type $N(S)=N_0\, S^\alpha $ to the cumulative source counts 
as a function of flux exceeding the threshold $S$ for both SZ-effects. In the table, values obtained from virial 
estimates are contrasted to values following from measurements on template data. The errors quoted denote the 95\% 
confidence intervals.}
\label{table_source_count_fits}
\end{table*}

The slopes derived from fits to the cluster number counts are slightly steeper for the virial estimates compared to 
template data ($\alpha\simeq -5/3$ versus $\alpha\simeq-1.4$), which again reflects the weaker dependence on cluster 
mass observed in template data. Comparing data sets for different frequencies, the slopes $\alpha$ are of course almost 
identical, because only amplitudes are changed by the choice of a different frequency band. The number of clusters $N_0$ 
stays roughly constant in the case of the kinetic SZ-effect, but reflects the distinct frequency modulation in the case 
of the thermal SZ-effect. Here, it should be emphasised, that the thermal SZ-effect does not vanish entirely at 
$\nu=217$~GHz due to \plancks symmetric frequency response window. The difference in numbers between the estimates 
based on virial quantities to those measured on template data amounts to roughly half an order of magnitude in the 
kinetic SZ-effect, but rises almost  an order of magnitude at small fluxes for the thermal SZ-effect. There is however 
good agreement in the number counts of thermal SZ clusters at high fluxes.

The difference in slope of the thermal versus kinetic SZ-cluster number counts is caused by the $M^{5/3}$-scaling of the 
thermal SZ-effect relative to the proportionality to $M$ of the kinetic effect. Due to the difference in slope, the 
effects are separated by two orders of magnitude for the largest fluxes, while this difference increases to eight orders 
of magnitude for the smallest fluxes, which hints at the difficulties to be faced in detecting kinetic clusters compared 
to even faint thermal detections. The slopes derived here are in good agreement with the those obtained by 
\citet{2001MNRAS.325..835K}

\section{Summary}\label{sect_summary}
All-sky maps for the thermal and kinetic Sunyaev-Zel'dovich effects are presented and their characteristics are 
described in detail. The maps because of their angular resolution and the data storage format chosen (HEALPix)
especially suited for simulations for \planck.

\begin{itemize}
\item{The all-sky maps of the thermal and kinetic SZ-effects presented here incorporate the correct 2-point correlation 
function, the evolving mass function and the correct size distribution of clusters, to within the accuracy of 
the underlying Hubble-volume simulation and the small-scale adiabatic gas simulations.}

\item{The maps presented here exhibit significant cluster substructure (compare Sect.~\ref{sect_skyview}). In spite of this, 
fits to the Comptonisation maps yield angular core radii, the distribution of which are close to the expectation based on the 
virial theorem (Sect.~\ref{sect_angsize}).}

\item{The difference in the distribution of the integrated Comptonisations $\mathcal{Y}$ and $\mathcal{W}$ 
(Sect.~\ref{sect_intwy}) and source counts $N(\bra S\ket>S_\mathrm{min})$ (Sect.~\ref{sect_sourcecount}) between values 
derived from scaling relations compared to those following from template data have been found to be substantial, which 
hints at possible misestimations of the number of clusters detectable for \planck.}

\item{An analytic investigation in Sect.~\ref{sect_szbg} quantified the contribution of the cluster sample to the 
sky averaged mean thermal Comptonisation $\bra y\ket$ and its variance $\sigma_y$. It was found that the 
clusters within the boundaries in mass ($M > 5\times 10^{13}M_{\sun}/h$) and redshift ($z < 1.48$) make up 
$\simeq40$\% of the mean Comptonisation, but account for $\sim98$\% of the variance. The value for the mean 
Comptonisation corresponds well to that obtained by other authors (Sect.~\ref{sect_pixamp}).}

\item{The power spectra (Sect.~\ref{sect_powerspec}) are compatible in amplitude and slope to the ones found by 
\citet{2002ApJ...579...16W}. On large angular scales, i.e. at small multipole $\ell$, deviations from the Poissonianity 
in the slope of the power spectrum have been found.}

\item{The velocities of the kinetic SZ-map correspond to the actual cosmological density environment, i.e. correlated infall 
velocities are observed due to the formation of superclusters, which highlights a significant improvement in comparison to 
methods that draw a cluster peculiar velocity from a (Gaussian) distribution and enables searches for the kinetic 
SZ-effect by considering spatial correlations with the thermal SZ-effect. The cross correlation of the thermal with the 
kinetic SZ-map yields a spectrum similar in shape at intermediate amplitudes (see Sect.~\ref{sect_powerspec}).}

\end{itemize}

Despite the high level of authenticity that the all-sky SZ-maps exhibit, there are quite a few issues not being taken 
account of: The baryon distribution and temperature inside the ICM is governed by processes beyond adiabatic gas 
physics, for example in the form of supernova feedback and radiative cooling. Especially the latter process gives rise to cool 
cores which may enhance the thermal SZ-signal. The ionisation inside the clusters was assumed to be complete. Furthermore, the 
maps contain only collapsed objects and hence filamentary structures or diffuse gas are not included. Concerning the thermal 
history of the ICM, reionisation had to be neglected. The kinetic map has been constructed without taking account of velocity 
fluctuations inside the cluster. This does not pose a problem for \planck, but needs to be remedied in high-resolution 
SZ-surveys to be undertaken with the {\em Atacama Cosmology Telescope} and the {\em South Pole Telescope}. Yet another 
imperfection is the lack of non-thermal particle populations that cause the relativistic SZ-effect \citep{2000A&A...360..417E}, 
whose detectability with \planck~ is still a matter of debate.


\section*{Acknowledgements}
The authors would like to thank Simon D.M. White for valuable comments.

\bibliography{bibtex/aamnem,bibtex/references}

\begin{thebibliography}{}

\bibitem[\protect\citeauthoryear{{Aghanim}, {de Luca}, {Bouchet}, {Gispert} \&
  {Puget}}{{Aghanim} et~al.}{1997}]{1997A&A...325....9A}
{Aghanim} N.,  {de Luca} A.,  {Bouchet} F.~R.,  {Gispert} R.,    {Puget} J.~L.,
   1997, \aap, 325, 9

\bibitem[\protect\citeauthoryear{{Arnaud} \& {Evrard}}{{Arnaud} \&
  {Evrard}}{1999}]{1999MNRAS.305..631A}
{Arnaud} M.,  {Evrard} A.~E.,  1999, \mnras, 305, 631

\bibitem[\protect\citeauthoryear{{Bartelmann}}{{Bartelmann}}{2001}]{2001A&A...%
370..754B}
{Bartelmann} M.,  2001, \aap, 370, 754

\bibitem[\protect\citeauthoryear{{Birkinshaw}}{{Birkinshaw}}{1999}]{1993birkin%
shaw}
{Birkinshaw} M.,  1999, Phys. Rep., 310, 98

\bibitem[\protect\citeauthoryear{{Bouchet} \& {Gispert}}{{Bouchet} \&
  {Gispert}}{1999}]{1999NewA....4..443B}
{Bouchet} F.~R.,  {Gispert} R.,  1999, New Astronomy, 4, 443

\bibitem[\protect\citeauthoryear{{Carlstrom}, {Holder} \& {Reese}}{{Carlstrom}
  et~al.}{2002}]{2002ARA&A..40..643C}
{Carlstrom} J.~E.,  {Holder} G.~P.,    {Reese} E.~D.,  2002, \araa, 40, 643

\bibitem[\protect\citeauthoryear{{Cavaliere} \& {Fusco-Femiano}}{{Cavaliere} \&
  {Fusco-Femiano}}{1978}]{1978A&A....70..677C}
{Cavaliere} A.,  {Fusco-Femiano} R.,  1978, \aap, 70, 677

\bibitem[\protect\citeauthoryear{{Colberg}, {White}, {Yoshida}, {MacFarland},
  {Jenkins}, {Frenk}, {Pearce}, {Evrard}, {Couchman}, {Efstathiou}, {Peacock},
  {Thomas} \& {The Virgo Consortium}}{{Colberg}
  et~al.}{2000}]{2000MNRAS.319..209C}
{Colberg} J.~M.,  {White} S.~D.~M.,  {Yoshida} N.,  {MacFarland} T.~J.,
  {Jenkins} A.,  {Frenk} C.~S.,  {Pearce} F.~R.,  {Evrard} A.~E.,  {Couchman}
  H.~M.~P.,  {Efstathiou} G.,  {Peacock} J.~A.,  {Thomas} P.~A.,    {The Virgo
  Consortium} 2000, \mnras, 319, 209

\bibitem[\protect\citeauthoryear{{da Silva}, {Kay}, {Liddle}, {Thomas},
  {Pearce} \& {Barbosa}}{{da Silva} et~al.}{2001}]{2001ApJ...561L..15D}
{da Silva} A.~C.,  {Kay} S.~T.,  {Liddle} A.~R.,  {Thomas} P.~A.,  {Pearce}
  F.~R.,    {Barbosa} D.,  2001, \apjl, 561, L15

\bibitem[\protect\citeauthoryear{{Delabrouille}, {Melin} \&
  {Bartlett}}{{Delabrouille} et~al.}{2002}]{2002hzcm.conf...81D}
{Delabrouille} J.,  {Melin} J.-B.,    {Bartlett} J.~G.,  2002, in ASP Conf.
  Ser. 257: AMiBA 2001: High-Z Clusters, Missing Baryons, and CMB Polarization
  {Simulations of Sunyaev-Zel'dovich Maps and Their Applications}.
pp 81--097

\bibitem[\protect\citeauthoryear{{Eke}, {Cole} \& {Frenk}}{{Eke}
  et~al.}{1996}]{1996MNRAS.282..263E}
{Eke} V.~R.,  {Cole} S.,    {Frenk} C.~S.,  1996, \mnras, 282, 263

\bibitem[\protect\citeauthoryear{{En{\ss}lin} \& {Kaiser}}{{En{\ss}lin} \&
  {Kaiser}}{2000}]{2000A&A...360..417E}
{En{\ss}lin} T.~A.,  {Kaiser} C.~R.,  2000, \aap, 360, 417

\bibitem[\protect\citeauthoryear{{Evrard}, {MacFarland}, {Couchman}, {Colberg},
  {Yoshida}, {White}, {Jenkins}, {Frenk}, {Pearce}, {Peacock} \&
  {Thomas}}{{Evrard} et~al.}{2002}]{2002ApJ...573....7E}
{Evrard} A.~E.,  {MacFarland} T.~J.,  {Couchman} H.~M.~P.,  {Colberg} J.~M.,
  {Yoshida} N.,  {White} S.~D.~M.,  {Jenkins} A.,  {Frenk} C.~S.,  {Pearce}
  F.~R.,  {Peacock} J.~A.,    {Thomas} P.~A.,  2002, \apj, 573, 7

\bibitem[\protect\citeauthoryear{{G{\' o}rski}, {Banday}, {Hivon} \&
  {Wandelt}}{{G{\' o}rski} et~al.}{2002}]{2002adass..11..107G}
{G{\' o}rski} K.~M.,  {Banday} A.~J.,  {Hivon} E.,    {Wandelt} B.~D.,  2002,
  in ASP Conf. Ser. 281: Astronomical Data Analysis Software and Systems XI
  {HEALPix --- a Framework for High Resolution, Fast Analysis on the Sphere}.
pp 107--+

\bibitem[\protect\citeauthoryear{{Herranz}, {Sanz}, {Hobson}, {Barreiro},
  {Diego}, {Mart{\'{\i}}nez-Gonz{\' a}lez} \& {Lasenby}}{{Herranz}
  et~al.}{2002}]{2002MNRAS.336.1057H}
{Herranz} D.,  {Sanz} J.~L.,  {Hobson} M.~P.,  {Barreiro} R.~B.,  {Diego}
  J.~M.,  {Mart{\'{\i}}nez-Gonz{\' a}lez} E.,    {Lasenby} A.~N.,  2002,
  \mnras, 336, 1057

\bibitem[\protect\citeauthoryear{{Hobson} \& {McLachlan}}{{Hobson} \&
  {McLachlan}}{2003}]{2003MNRAS.338..765H}
{Hobson} M.~P.,  {McLachlan} C.,  2003, \mnras, 338, 765

\bibitem[\protect\citeauthoryear{{Jenkins}, {Frenk}, {White}, {Colberg},
  {Cole}, {Evrard}, {Couchman} \& {Yoshida}}{{Jenkins}
  et~al.}{2001}]{2001MNRAS.321..372J}
{Jenkins} A.,  {Frenk} C.~S.,  {White} S.~D.~M.,  {Colberg} J.~M.,  {Cole} S.,
  {Evrard} A.~E.,  {Couchman} H.~M.~P.,    {Yoshida} N.,  2001, \mnras, 321,
  372

\bibitem[\protect\citeauthoryear{{Kay}, {Liddle} \& {Thomas}}{{Kay}
  et~al.}{2001}]{2001MNRAS.325..835K}
{Kay} S.~T.,  {Liddle} A.~R.,    {Thomas} P.~A.,  2001, \mnras, 325, 835

\bibitem[\protect\citeauthoryear{{Lloyd-Davies}, {Ponman} \&
  {Cannon}}{{Lloyd-Davies} et~al.}{2000}]{2000MNRAS.315..689L}
{Lloyd-Davies} E.~J.,  {Ponman} T.~J.,    {Cannon} D.~B.,  2000, \mnras, 315,
  689

\bibitem[\protect\citeauthoryear{{Mandolesi}, {Bersanelli}, {Cesarsky},
  {Danese}, {Efstathiou}, {Griffin}, {Lamarre}, {Norgaard-Nielsen}, {Pace},
  {Puget}, {Raisanen}, {Smoot}, {Tauber} \& {Volonte}}{{Mandolesi}
  et~al.}{1995}]{1995P&SS...43.1459M}
{Mandolesi} N.,  {Bersanelli} M.,  {Cesarsky} C.,  {Danese} L.,  {Efstathiou}
  G.,  {Griffin} M.,  {Lamarre} J.~M.,  {Norgaard-Nielsen} H.~U.,  {Pace} O.,
  {Puget} J.~L.,  {Raisanen} A.,  {Smoot} G.~F.,  {Tauber} J.,    {Volonte} S.,
   1995, \planss, 43, 1459

\bibitem[\protect\citeauthoryear{{Mathiesen} \& {Evrard}}{{Mathiesen} \&
  {Evrard}}{2001}]{2001ApJ...546..100M}
{Mathiesen} B.~F.,  {Evrard} A.~E.,  2001, \apj, 546, 100

\bibitem[\protect\citeauthoryear{{Mohr}, {Mathiesen} \& {Evrard}}{{Mohr}
  et~al.}{1999}]{1999ApJ...517..627M}
{Mohr} J.~J.,  {Mathiesen} B.,    {Evrard} A.~E.,  1999, \apj, 517, 627

\bibitem[\protect\citeauthoryear{{Monaghan} \& {Lattanzio}}{{Monaghan} \&
  {Lattanzio}}{1985}]{1985A&A...149..135M}
{Monaghan} J.~J.,  {Lattanzio} J.~C.,  1985, \aap, 149, 135

\bibitem[\protect\citeauthoryear{{Moscardini}, {Bartelmann}, {Matarrese} \&
  {Andreani}}{{Moscardini} et~al.}{2002}]{2002MNRAS.335..984M}
{Moscardini} L.,  {Bartelmann} M.,  {Matarrese} S.,    {Andreani} P.,  2002,
  \mnras, 335, 984

\bibitem[\protect\citeauthoryear{{Myers}, {Shanks}, {Outram}, {Frith} \&
  {Wolfendale}}{{Myers} et~al.}{2004}]{2004MNRAS.347L..67M}
{Myers} A.~D.,  {Shanks} T.,  {Outram} P.~J.,  {Frith} W.~J.,    {Wolfendale}
  A.~W.,  2004, \mnras, 347, L67

\bibitem[\protect\citeauthoryear{{Nagai}, {Kravtsov} \& {Kosowsky}}{{Nagai}
  et~al.}{2003}]{2003ApJ...587..524N}
{Nagai} D.,  {Kravtsov} A.~V.,    {Kosowsky} A.,  2003, \apj, 587, 524

\bibitem[\protect\citeauthoryear{{Navarro}, {Frenk} \& {White}}{{Navarro}
  et~al.}{1995}]{1995MNRAS.275..720N}
{Navarro} J.~F.,  {Frenk} C.~S.,    {White} S.~D.~M.,  1995, \mnras, 275, 720

\bibitem[\protect\citeauthoryear{{Persi}, {Spergel}, {Cen} \&
  {Ostriker}}{{Persi} et~al.}{1995}]{1995ApJ...442....1P}
{Persi} F.~M.,  {Spergel} D.~N.,  {Cen} R.,    {Ostriker} J.~P.,  1995, \apj,
  442, 1

\bibitem[\protect\citeauthoryear{{Press} \& {Schechter}}{{Press} \&
  {Schechter}}{1974}]{1974ApJ...187..425P}
{Press} W.~H.,  {Schechter} P.,  1974, \apj, 187, 425

\bibitem[\protect\citeauthoryear{{Refregier}, {Komatsu}, {Spergel} \&
  {Pen}}{{Refregier} et~al.}{2000}]{2000PhRvD..61l3001R}
{Refregier} A.,  {Komatsu} E.,  {Spergel} D.~N.,    {Pen} U.,  2000, \prd, 61,
  123001

\bibitem[\protect\citeauthoryear{{Rephaeli}}{{Rephaeli}}{1995}]{1995ARA&A..33.%
.541R}
{Rephaeli} Y.,  1995, \araa, 33, 541

\bibitem[\protect\citeauthoryear{{Scaramella}, {Cen} \&
  {Ostriker}}{{Scaramella} et~al.}{1993}]{1993ApJ...416..399S}
{Scaramella} R.,  {Cen} R.,    {Ostriker} J.~P.,  1993, \apj, 416, 399

\bibitem[\protect\citeauthoryear{{Seljak}, {Burwell} \& {Pen}}{{Seljak}
  et~al.}{2001}]{2001PhRvD..63f3001S}
{Seljak} U.,  {Burwell} J.,    {Pen} U.,  2001, \prd, 63, 063001

\bibitem[\protect\citeauthoryear{{Sheth} \& {Tormen}}{{Sheth} \&
  {Tormen}}{1999}]{1999MNRAS.308..119S}
{Sheth} R.~K.,  {Tormen} G.,  1999, \mnras, 308, 119

\bibitem[\protect\citeauthoryear{{Spergel}, {Verde}, {Peiris}, {Komatsu},
  {Nolta}, {Bennett}, {Halpern}, {Hinshaw}, {Jarosik}, {Kogut}, {Limon},
  {Meyer}, {Page}, {Tucker}, {Weiland}, {Wollack} \& {Wright}}{{Spergel}
  et~al.}{2003}]{2003astro.ph..2209S}
{Spergel} D.~N.,  {Verde} L.,  {Peiris} H.~V.,  {Komatsu} E.,  {Nolta} M.~R.,
  {Bennett} C.~L.,  {Halpern} M.,  {Hinshaw} G.,  {Jarosik} N.,  {Kogut} A.,
  {Limon} M.,  {Meyer} S.~S.,  {Page} L.,  {Tucker} G.~S.,  {Weiland} J.~L.,
  {Wollack} E.,    {Wright} E.~L.,  2003, \apj, submitted, astro-ph/0302209

\bibitem[\protect\citeauthoryear{{Springel} \& {Hernquist}}{{Springel} \&
  {Hernquist}}{2002}]{2002MNRAS.333..649S}
{Springel} V.,  {Hernquist} L.,  2002, \mnras, 333, 649

\bibitem[\protect\citeauthoryear{{Springel}, {Yoshida} \& {White}}{{Springel}
  et~al.}{2001}]{2001NewA....6...79S}
{Springel} V.,  {Yoshida} N.,    {White} S.~D.~M.,  2001, New Astronomy, 6, 79

\bibitem[\protect\citeauthoryear{{Stoehr}}{{Stoehr}}{1999}]{FelixStoehr1999}
{Stoehr} F., , 1999, High Resolution Simulations of Underdense Regions, Diploma
  Thesis, Munich: Technical University

\bibitem[\protect\citeauthoryear{{Sunyaev} \& {Zel'dovich}}{{Sunyaev} \&
  {Zel'dovich}}{1972}]{1972SZorig}
{Sunyaev} R.~A.,  {Zel'dovich} I.~B.,  1972, Comments Astrophys. Space Phys.,
  4, 173

\bibitem[\protect\citeauthoryear{{Sunyaev} \& {Zel'dovich}}{{Sunyaev} \&
  {Zel'dovich}}{1980}]{1980ARA&A..18..537S}
{Sunyaev} R.~A.,  {Zel'dovich} I.~B.,  1980, \araa, 18, 537

\bibitem[\protect\citeauthoryear{{Tauber}}{{Tauber}}{2000}]{2000IAUS..201E...8%
T}
{Tauber} J.~A.,  2000, in IAU Symposium {The Planck Mission}

\bibitem[\protect\citeauthoryear{{White}}{{White}}{2003}]{2003ApJ...597..650W}
{White} M.,  2003, \apj, 597, 650

\bibitem[\protect\citeauthoryear{{White}, {Hernquist} \& {Springel}}{{White}
  et~al.}{2002}]{2002ApJ...579...16W}
{White} M.,  {Hernquist} L.,    {Springel} V.,  2002, \apj, 579, 16

\bibitem[\protect\citeauthoryear{{Wright}}{{Wright}}{1979}]{1979ApJ...232..348%
W}
{Wright} E.~L.,  1979, \apj, 232, 348

\end{thebibliography}
\bibliographystyle{mn2e}

\appendix

\bsp

\label{lastpage}

\end{document}